\RequirePackage{fix-cm}

\documentclass[smallcondensed]{svjour3}
\smartqed  

\usepackage{amsmath, latexsym, amssymb}
\usepackage[plain]{algorithm2e}
\usepackage[colorlinks=true,citecolor=blue,linkcolor=blue,urlcolor=blue]{hyperref}
\usepackage{graphicx}
\usepackage{subcaption}

\newcommand{\Rmnum}[1]{\expandafter\@slowromancap\romannumeral #1@}

\def\mvp{\vspace*{-0.2in}}

\newcommand{\mund}[1]{\noindent{\bf #1}}
\newcommand{\Ignore}[1]{}

\begin{document}
\title{Bayes Networks for Supporting Query Processing Over Incomplete Autonomous Databases
	\thanks{This research is supported by ONR grant N000140910032
	and two Google research awards.}}

\titlerunning{Bayes Nets for Query Processing Over Incomplete Autonomous Databases}  

\author{Rohit Raghunathan \and
		Sushovan De         \and
        Subbarao Kambhampati
}

\institute{R. Raghunathan \at
              Amazon \\
              \email{rraghun@amazon.com}
			  \and
S. De \and S. Kambhampati \at
              Arizona State University \\
              Computer Science and Engineering\\
              Tel: +1 (480) 965-2735\\
              \email{\{sushovan, rao\} @asu.edu}
}

\date{}

\maketitle

\begin{abstract}
As the information available to lay users through autonomous data
sources continues to increase, mediators become important to ensure
that the wealth of information available is tapped effectively. A key
challenge that these information mediators need to handle is the
varying levels of incompleteness in the underlying databases in terms
of missing attribute values. Existing approaches such as QPIAD aim to mine
and use Approximate Functional Dependencies (AFDs) to predict and
retrieve relevant incomplete tuples. These approaches make
independence assumptions about missing values---which critically
hobbles their performance when there are tuples containing missing
values for multiple correlated attributes. In this paper, we present a
principled probabilistic alternative that views an incomplete tuple as
defining a distribution over the complete tuples that it stands
for. We learn this distribution in terms of Bayes networks. Our
approach involves mining/``learning" Bayes networks from a sample of
the database, and using it to do both imputation (predict a missing
value) and query rewriting (retrieve relevant results with
incompleteness on the query-constrained attributes, when the data
sources are autonomous). We present empirical studies to demonstrate
that (i) at higher levels of incompleteness, when multiple attribute
values are missing, Bayes networks do provide a significantly higher
classification accuracy and (ii) the relevant possible answers
retrieved by the queries reformulated using Bayes networks provide
higher precision and recall than AFDs while keeping query processing
costs manageable.
\end{abstract}
 

\setcounter{secnumdepth}{3}
\section{Introduction} 

As the popularity of the World Wide Web continues to increase, lay
users have access to more and more information in autonomous
databases. Incompleteness in these autonomous sources is extremely
commonplace. Such incompleteness mainly arises due to the way in which
these databases are populated --- by lay users, or through (inaccurate) 
automatic extraction. Dealing with incompleteness in the
databases requires tools for dealing with uncertainty. Previous
attempts at dealing with this uncertainty by systems like
QPIAD~\cite{qpiad-paper} have mainly focused on using rule-based
approaches, popularly known in the database community as Approximate
Functional Dependencies (AFDs). The appeal of AFDs is due to the ease
of specifying the dependencies, learning and reasoning with
uncertainty. However, uncertain reasoning using AFDs adopts the
certainty factors model, which assumes that the principles of locality
and detachment~\cite{aima-book} hold. But, these principles do not
hold for uncertain reasoning and can lead to erroneous
reasoning. \Ignore{To make rule-based systems useful in practice,
  early expert systems like the MYCIN medical diagnosis program used
  certainty factors model which allowed rule sets either in the causal
  or in the diagnostic direction only, limiting the applicability of
  the system.} As the levels of incompleteness in the information
sources increases, the need for more scalable and accurate reasoning
becomes paramount. 

Full probabilistic reasoning avoids the traps of
AFDs. Graphical models are an efficient way of doing full
probabilistic reasoning. Bayesian network (Bayes net) is such a model,
where direct dependencies between the variables in a problem are
modeled as a directed acyclic graph, and the indirect dependencies can
be inferred. As desired, Bayes nets can model both causal and
diagnostic dependencies.
Using Bayes nets for uncertain reasoning has largely replaced rule-based approaches in Artificial Intelligence. However, learning and inference on Bayes nets can be computationally expensive which might inhibit their applications to handling incompleteness in autonomous data sources. In this paper, we consider if these costs can be handled without compromising on the improved accuracy offered by Bayes nets, in the context of incompleteness in the autonomous databases.

\medskip
\mund{Incompleteness in Autonomous Databases:}
Increasingly many of the autonomous web databases are being populated
by automated techniques or by lay users, with very little
curation. For example, databases like autotrader.com are populated
using automated extraction techniques by crawling the text classifieds
and by car owners entering data through forms. Scientific databases
such as CbioC~\cite{cbioc-website}, also use similar techniques for
populating the database. However, Gupta and
Sarawagi~\cite{incorrect-extraction-paper} have shown that these
techniques are error prone and lead to incompleteness in the database
in the sense that many of the attributes have missing values. Wolf et
al~\cite{qpiad-paper} report that 99\% of the 35,000 tuples extracted
from Cars Direct were incomplete.  When the mediator has privileges to
modify the data sources, the missing values in these data sources can
be completed using "imputation", which attempts to fill in the missing
values with the most likely value. As the levels of incompleteness in
these data sources increase, it is not uncommon to come across tuples
with multiple missing values. Effectively finding the most likely
completions for these multiple missing values would require capturing
the dependencies between them.  A second challenge arises when the
underlying data sources are autonomous, i.e., access to these
databases are through forms, the mediator cannot complete the missing
values with the most likely values. Therefore, mediators need to
generate and issue a set of reformulated queries, in order to retrieve
the relevant answers with missing values. Efficiency considerations
dictate that the number of reformulations be kept low. In such
scenarios, it becomes very important for mediators to send queries
that not only retrieve results with a large fraction of relevant
results (precision), but also a large number of relevant results
(recall).

\medskip
\mund{QPIAD \& AFDs:}
The QPIAD system~\cite{qpiad-paper} addresses the challenges in
retrieving relevant incomplete answers by learning the correlations
between the attributes in the database as AFDs and the value
distributions as Na\"ive Bayesian Classifiers.
AFDs are rule-based methods for dealing with uncertainty. AFDs adopt
the certainty factors model which makes two strong
assumptions:

\smallskip \textbf{1. Principle of Locality:} Whenever there
is a rule $A \rightarrow B$, given evidence of $A$, we can conclude $B$,
regardless of the other rules and evidences.

\smallskip
\textbf{2. Principle of Detachment:} Whenever a proposition B is found
to be true, the truth of B can be used regardless of how it was found
to be true.

\smallskip However, these two assumptions do not hold in the
presence of uncertainty. When propagating beliefs, not only is it
important to consider \emph{all} the evidences but also their
sources. Therefore, using AFDs for reasoning with uncertainty can lead
to cyclic reasoning and fail to capture the correlations between
multiple missing values.  In addition to these shortcomings, the
beliefs are represented using a Naive-Bayesian Classifier, which makes
strong conditional independence assumptions, often leading to
inaccurate values.

\medskip
\mund{Overview of our Approach:}
Given the advantages of Bayes networks over AFDs, we investigate if
replacing AFDs with  Bayes networks in QPIAD
system~\cite{qpiad-paper}, provides higher accuracy and while keeping
the costs manageable. Learning and inference with Bayes networks are
computationally harder than AFDs. Therefore, the challenges involved
in replacing AFDs with Bayes networks include learning and using them
to do both imputation and query rewriting by keeping costs
manageable. We use BANJO software package~\cite{banjo} to learn the
topology of the Bayes network and use BNT~\cite{bnt} and
INFER.NET~\cite{infernet} software packages to do inference on them.
Even though learning the topology for the Bayes net from a sample of
the database involves searching over the possible topologies, we found
that high fidelity Bayes networks could be learnt from a small
fraction of the database by keeping costs manageable (in terms of time
spent in searching).  Inference in Bayes networks is intractable if
the network is multiply connected, i.e., there is more than one undirected
path between any two nodes in the network. We handle this challenge by
using approximate inference techniques. Approximate inference
techniques are able to retain the accuracy of exact inference
techniques and keep the cost of inference manageable. We compare the
cost and accuracy of using AFDs and Bayes networks for imputing single
and multiple missing values at different levels of incompleteness in
test data.

We also develop new techniques for generating rewritten queries using
Bayes networks. 
The three challenges that are involved
in generating rewritten queries are:
\smallskip 1. Selecting the
attributes on which the new queries will be formulated. Selecting
these attributes by searching over all the attributes becomes too
expensive as the number of attributes in the database
increases.
\smallskip 2. Determining the values to which the attributes
in the rewritten query will be constrained to. The size of the domains
of attributes in most autonomous databases is often large. Searching
over each and every value can be expensive.
\smallskip 3. Most autonomous
data sources have a limit on the number of queries to which it will
answer. The rewritten queries that we generate should be able to
carefully tradeoff precision with the throughput of the results
returned.
\smallskip We propose techniques to handle these challenges and
compare them with AFD-based approaches in terms of precision and
recall of the results returned.

\medskip\mund{Organization:}
The rest of the paper is organized as follows --- We begin with a discussion of related work, then in Section~\ref{sec:problem-setting}, we describe the problem setting and background. In Section~\ref{ch:learning-and-imputation}, we discuss how Bayes network models of autonomous databases can be learnt by keeping costs manageable. In Section~\ref{sec:imputation}, we compare the prediction accuracy and cost of using Bayes network and AFDs for imputing missing values. Next, in Section~\ref{sec:rewriting}, we discuss how rewritten queries are generated using Bayes networks and compare them with AFD-approaches for single and multi-attribute queries. Finally, we conclude in Section~\ref{sec:conclusion}.

\section{Related Work}
\label{sec:related}
This work is a significant extension of the QPIAD system \cite{qpiad-paper,wolf2007query}, which also deals with incompleteness in databases. While the QPIAD system also learns attribute correlations, it does so using Approximate Functional Dependencies (AFDs) and uses Naive Bayesian Classifiers for representing value distributions and reformulating queries. Additionally, the QPIAD system can only handle missing values on a single attribute. In contrast, we use Bayes Network models learned from a sample of the database to represent attribute correlations and value distributions. We use the methods used in the QPIAD system as our baseline approach. 

Completing missing values in databases using Bayesian Networks has been addressed previously \cite{batista2002study,dempster1977maximum,little1987statistical,ramoni2001robust,wu2004using}. But most methods focus on completing the missing values so as to preserve the original data statistics so that other data mining techniques can be applied to it. We concentrate on retrieving relevant possible answers in the presence of missing values on single and multiple attributes. For example, \cite{wu2004using} uses association rules to impute the value of the missing attributes, whereas we use Bayes Networks to impute the value as well as retrieve results.

Like QPIAD, and other work on querying over incomplete databases, we too assume that the level of incompleteness in the database is small enough that it is possible to get a training sample that is mostly complete. Thus, we use techniques for learning from complete training data. If the training sample itself were to be incomplete, then we will need to employ expectation-maximization techniques during learning \cite{dempster1977maximum,ramoni1997learning}. 


Work on querying inconsistent databases usually focuses on fixing problems with the query itself \cite{muslea2005online,nambiar2006answering}. If the query has an empty resultset, or if the query does not include all relevant keywords, it can be automatically augmented to fix those shortcomings. The objective of this work is to deal with shortcomings of the data --- our query rewriting algorithms help retrieve useful tuples even in the presence of multiple missing values in them. 

\section{Problem Setting \& Background}
\label{sec:problem-setting}

\mund{Overview of QPIAD:}
Since our main comparison is with the QPIAD system \cite{qpiad-paper}, we will provide a
brief overview of its operation. 
Given a Relation \emph{R}, a subset \emph{X} of its attributes, and a single attribute \emph{A} of \emph{R}, an approximate functional dependency (AFD) holds on a Relation \emph{R}, between \emph{X} and \emph{A}, denoted by, \emph{X}~$\leadsto$~\emph{A}, if the corresponding functional dependency \emph{X}~$\rightarrow$~\emph{A} holds on all but a small fraction of tuples of \emph{R}.

To illustrate how QPIAD works consider the the query $\text{\emph{Q : Body=SUV}}$ issued to Table~\ref{Fragment-Cars-DB-challenge-section}. Traditional query processors will only retrieve tuples $\text{\emph{t}}_{\text{\emph{7}}}$ and $\text{\emph{t}}_{\text{\emph{9}}}$.  However, the entities represented by tuples $\text{\emph{t}}_{\text{\emph{8}}}$ and $\text{\emph{t}}_{\text{\emph{10}}}$ are also likely to be relevant. The QPIAD system's aim is to retrieve tuples $\text{\emph{t}}_{\text{\emph{8}}}$ and $\text{\emph{t}}_{\text{\emph{10}}}$, in addition to $\text{\emph{t}}_{\text{\emph{7}}}$ and $\text{\emph{t}}_{\text{\emph{9}}}$. 
In order to retrieve tuples $\text{\emph{t}}_{\text{\emph{8}}}$ and $\text{\emph{t}}_{\text{\emph{10}}}$ it uses AFDs mined from a sample of the database. For example, an AFD \emph{Model}~$\leadsto$~\emph{Body} may be mined for the fragment of the cars database shown in Table~\ref{Fragment-Cars-DB-challenge-section}. This indicates that the value of a car's \emph{Model} attribute \emph{often} (but not always) determines the value of its \emph{Body} attribute. These rules are used to retrieve relevant incomplete answers.

\begin{table}
{\footnotesize
\begin{center}
\begin{tabular}{ | p{0.5cm} | p{1.2cm} | p{1cm} | p{1cm} | p{1cm}| p{1cm}| p{1cm}|}

    \hline
    ID & Make    & Model & Year & Body  & Mileage \\ \hline\hline
    1  & Audi    & null  & null & Sedan & 20000   \\ \hline
    2  & Audi    & A8    & null & Sedan & 15000   \\ \hline
    3  & BMW     & 745   & 2002 & Sedan & 40000   \\ \hline
    4  & Audi    & null  & 2005 & Sedan & 20000   \\ \hline
	5  & Audi    & A8    & 2005 & Sedan & 20000   \\ \hline
    6  & BMW     & 645   & 1999 & Convt & null    \\ \hline
    7  & Hyundai & Santa & 1990 & SUV   & 45000   \\ \hline
    8  & Hyundai & Santa & 1993 & null  & 40000   \\ \hline
    9  & Acura   & MDX   & 1990 & SUV   & 30000   \\ \hline
    10 & Acura   & MDX   & 1990 & null  & 12000   \\ \hline
    \end{tabular}
\end{center}
}
\mvp\caption{A Fragment of a car database.\label{Fragment-Cars-DB-challenge-section}}
\end{table}

When the mediators have access privileges to modify the database, AFDs are used along with Naive Bayes Classifiers to fill in the missing values as a simple classification task and then traditional query processing will suffice to retrieve relevant answers with missing values. However, in more realistic scenarios, when such privileges are not provided, mediators generate a set of rewritten queries and send to the database, in addition to the original user query. 
According to the AFD mentioned above and tuple $\text{\emph{t}}_{\text{\emph{7}}}$ retrieved by traditional query processors, a rewritten query $\text{\emph{Q'}}_{\text{\emph{1}}}:\sigma_{\text{\emph{Model=Santa}}}$ may be generated to retrieve $\text{\emph{t}}_{\text{\emph{8}}}$. Similarly $\text{\emph{Q'}}_{\text{\emph{2}}}:\sigma_{\text{\emph{Model=MDX}}}$ may be generated which will retrieve $\text{\emph{t}}_{\text{\emph{10}}}$.

Multiple rules can be mined for each attribute, for example, the mileage and year of a car might determine the body style of the car. So a rule \emph{\{Year, Mileage\}} $\leadsto$ \emph{\{Body\}} could be mined. Each rule has a confidence associated with it, which specifies how accurate the determining set of an attribute's AFD is in predicting it. The current QPIAD system uses only the highest confidence AFD\footnote[1]{The actual implementation of QPIAD uses a variant to the highest confidence AFD for some of the attributes. For details we refer the reader to~\cite{qpiad-paper}} of each attribute for imputation and query rewriting. In addition, it only aims to retrieve relevant incomplete answers with atmost one missing value on query-constrained attributes.


%
To illustrate the shortcomings of AFDs, consider a query \emph{Q} : $\sigma_{\text{\emph{Model=A8}}\wedge\text{\emph{Year=2005}}}$ issued to the fragment of the car database shown in Table~\ref{Fragment-Cars-DB-challenge-section}. When the mediator has modification privileges, the missing values for attributes \emph{Model} and \emph{Year} can be completed with the most likely values, before returning the answer set. Using AFDs to predict the missing values in tuple $\text{\emph{t}}_{\text{\emph{1}}}$, ignores the correlation between the \emph{Model} and \emph{Year}; predicting them independently. Substituting the value for missing attribute \emph{Year} in tuple $\text{\emph{t}}_{\text{\emph{2}}}$ using just the highest confidence rule as is done in QPIAD~\cite{qpiad-paper}, often leads to inaccurate propagation of beliefs as the other rules are ignored.
When the mediator does not have privileges to modify the database, a set of rewritten queries are generated and issued to the database to retrieve the relevant uncertain answers. Issuing \emph{Q} to the database fragment in Table~\ref{Fragment-Cars-DB-challenge-section} retrieves $\text{\emph{t}}_{\text{\emph{5}}}$. The rewritten queries generated by methods discussed in QPIAD~\cite{qpiad-paper} retrieve tuples $\text{\emph{t}}_{\text{\emph{2}}}$ and $\text{\emph{t}}_{\text{\emph{4}}}$. However, it does not retrieve tuple $\text{\emph{t}}_{\text{\emph{1}}}$, but it is highly possible that the entity represented by it is relevant to the user's query.

\medskip\mund{Bayes Networks:}
A Bayes network~\cite{pgm-book} is a graphical representation of the probabilistic dependencies between the variables in a domain. The generative model of a relational database can be represented using a Bayes network, where each node in the network represents an attribute in the database. 
The edges between the nodes represent direct probabilistic dependencies between the attributes. The strength of these probabilistic dependencies are modeled by associating a conditional probability distribution (CPD) with each node, which represents the conditional probability of a variable, given as evidence each combination of values of its immediate parents.
A Bayes network is a compact representation of the full joint probability distribution of the nodes in the network. The full joint distribution can be constructed from the CPDs in the Bayes network. Given the full joint distribution, any probabilistic query can be answered. In particular, the probability of any set of hypotheses can be computed, given any set of observations, by conditioning and marginalizing over the joint distribution. 
Since the semantics of Bayes networks are in terms of the full joint probability distribution, inference using them considers the influence of all variables in the network. Therefore, Bayes nets, unlike AFDs, do not make the Locality and Detachment assumptions.

\setcounter{secnumdepth}{3}
\section{Learning Bayes Network Models}
\label{ch:learning-and-imputation}

\begin{table*}[t]

\begin{center}
\begin{tabular}{ | p{1.8cm} | p{0.8cm} | p{0.8cm} | p{0.8cm} | p{0.8cm} | p{0.9cm} | p{0.8cm} | }

    \hline
    Database & Year & Model & Make & Price & Mileage & Body \\ \hline \hline
  
 Cars-8000-20(Mediator) & \centering 9 & \centering 38 & \centering 6 & \centering 19 & \centering 17 &  5 \\ \hline
 Cars-8000-100(Complete) & \centering 12 & \centering 41 & \centering 6 & \centering 30 & \centering 20 & 7\\ \hline
  \end{tabular}
\caption[Domain size of attributes in Car database]{Domain size of attributes in Car database.\label{tb:domain-size-cars}}
\end{center}
\end{table*}

\begin{table*}
\begin{center}

\begin{tabular}{ | p{1.8cm} | p{0.25cm} | p{0.5cm} | p{0.4cm} | p{0.8cm} | p{0.60cm} | p{1.0cm} |  p{0.45cm} |  p{0.4cm} |  p{0.55cm} |  p{1.0cm} |  }

    \hline
    Database & Age & Work Class & Educ-ation & Marital Status & Occu-pation & Relation-ship & Race & Sex & Hours Per Week & Native Country \\ \hline\hline
  
 Adult-15000-20(Mediator) & \centering 8 & \centering 7 & \centering 16 & \centering 7 & \centering 14 & \centering 6 &  \centering 5 & \centering  2 & \centering 10 &  37 \\ \hline
 Adult-15000-100(Complete)  & \centering 8 & \centering 7 & \centering 16 & \centering 7 & \centering 14 & \centering 6 & \centering 5 & \centering 2 & \centering 10 &  40\\ \hline

    \end{tabular}
\caption[Domain size of attributes in Adult database]{Domain size of attributes in Adult database.\label{tb:domain-size-adult}}
\end{center}
\end{table*}

\label{sec:bayes-net-learning}
In this section we discuss how we learn the topology and parameters of the Bayes network by keeping costs manageable. We learn the generative model of two databases --- A fragment of 8000 tuples extracted from \emph{Cars.com}~\cite{cars.com} and Adult database consisting of 15000 tuples obtained from UCI data repository~\cite{uci-repository}. Table~\ref{tb:domain-size-cars}~and~\ref{tb:domain-size-adult} describe the schema and the domain sizes of the attributes in the two databases. The attributes with continuous values are discretized and used as categorical attributes. \emph{Price} and \emph{Mileage} attributes in the cars database are discretized by rounding off to the nearest five thousand. In the adult database attributes \emph{Age} and \emph{Hours Per Week} are discretized to the nearest multiple of five. 

\begin{figure}
\begin{center}
	\begin{subfigure}{0.48\textwidth}
		\includegraphics[keepaspectratio, width=190pt, clip, trim= 50pt 250pt 0pt 50pt]{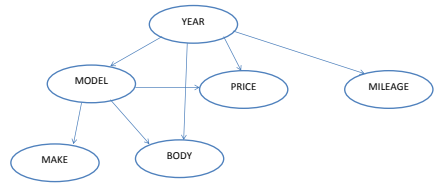}\\
		\caption{\emph{Cars.com} dataset}
		\vspace{0.2in}
		\label{example-bayes-network}
	\end{subfigure}

	\begin{subfigure}{0.48\textwidth}
		\includegraphics[keepaspectratio, width=190pt, clip, trim= 50pt 0pt 0pt 50pt]{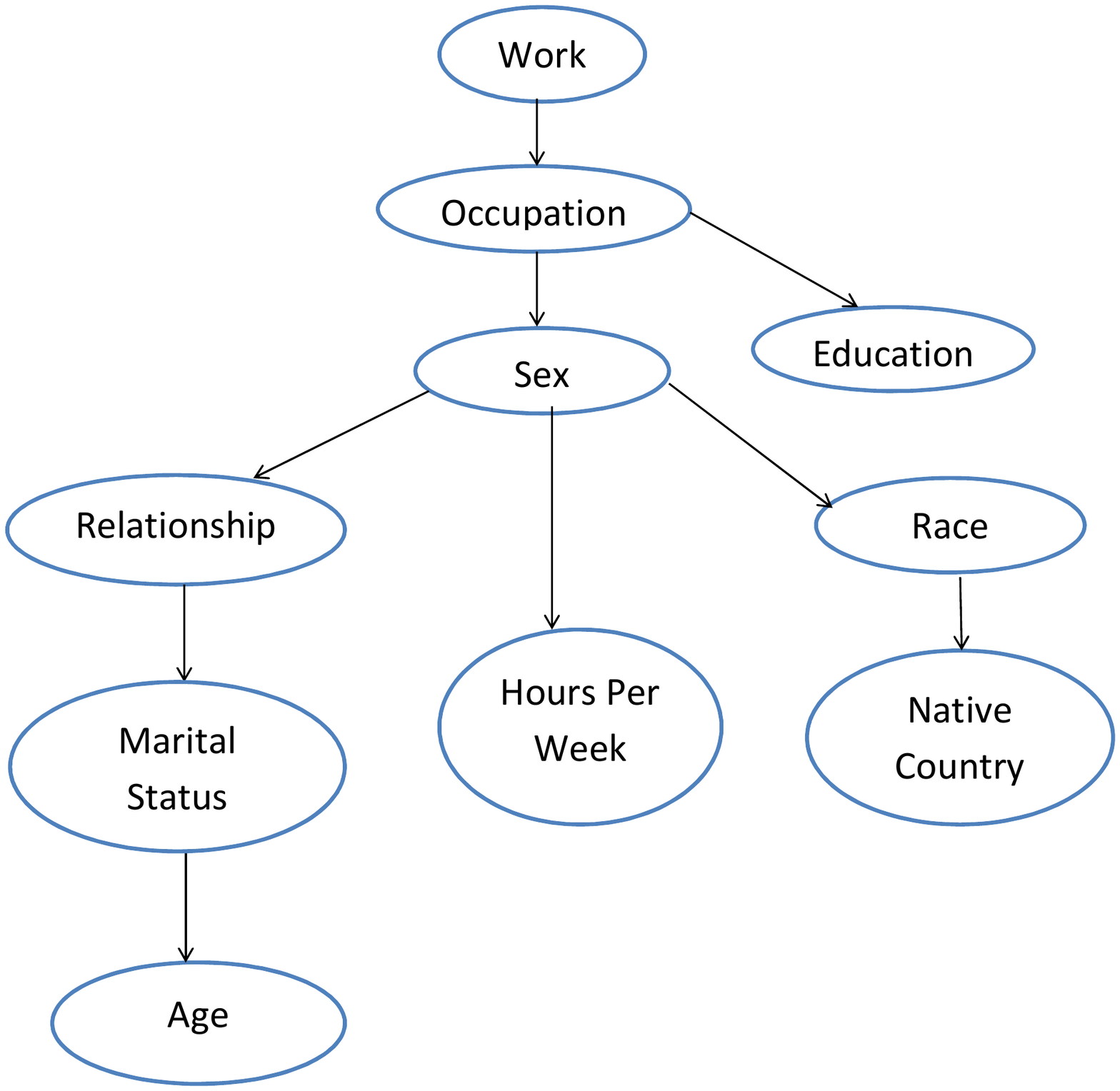}\\
		\vspace{-0.2in}
		\caption{Adult dataset}
		\label{fig:adult-bayes-network}
	\end{subfigure}
	\caption{Bayesian networks learned from a sample of the data}
\end{center}
\end{figure}

The structure of the Bayes network is learned from a complete sample of the autonomous database. We use the BANJO package~\cite{banjo} as a black box for learning the structure of the Bayes network. To keep the learning costs manageable we constrain nodes to have at most two parents. In cases where there are more than two attributes directly correlated to an attribute, these attributes can be modeled as children. There is no limit on the number of children a node can have. Figure~\ref{example-bayes-network} shows the structure of a Bayes network learned for the Cars database and Figure~\ref{fig:adult-bayes-network} for the Adult database. We used samples of sizes varying from 5-20\% of the database and found that the structure of the highest scoring network remained the same. We also experimented with different time limits for the search, ranging from 5-30 minutes. We did not see any change in the structure of the highest confidence network.

\Ignore{
\subsubsection{Parameter Learning}
\label{subsec:parameter-learning}
Learning the Maximum Likelihood parameters for the previously learnt Bayesian network structure when there is complete data is relatively cheap as the parameters are un-entagled. Therefore it does not involve any inference and the parameters can be learnt through data counts. When sufficient complete training data is not available, Bayes nets can learn the parameters from incomplete data using Expectation Maximization (EM) techniques. EM learns the parameters iteratively by performing inference and optimizations. We found that learning the parameters using a single iteration of EM gives the prediction accuracy of Bayes networks a distinct edge over AFDs. In section~\ref{subsec:impt-training} we discuss the costs and accuracy of using EM for learning parameters from incomplete training data.     
} 

Bayes network inference is used in both imputation and query rewriting tasks. Imputation involves substituting the missing values with the most likely values, which involves inference. Exact inference in Bayes networks is NP-hard~\cite{aima-book} if the network is multiply connected. Therefore, to keep query processing costs manageable we use approximate inference techniques. In our experiments described in section~\ref{sec:imputation}, we found that using approximate inference techniques retains the accuracy edge of exact inference techniques, while keeping the prediction costs manageable. We use the BNT package~\cite{bnt} for doing inference on the Bayes Network for the imputation task. We experimented with various exact inference engines that BNT offers and found the junction-tree engine to be the fastest. While querying multiple variables, junction tree inference engine can be used only when all the variables being queried form a clique. When this is not the case, we use the variable elimination inference engine. 

\section{Imputation using Bayes Networks}
\label{sec:imputation}
In this section we compare the prediction accuracy and cost of Bayes networks versus AFDs for imputing single and multiple missing values when there is incompleteness in test data. 
When the mediator has privileges to modify the underlying Autonomous database, the missing values can be substituted with the most probable value. Imputation using Bayesian networks first computes the posterior of the attribute that is to be predicted given the values present in the tuple and completes the missing value with the most likely value given the evidence. When predicting multiple missing values, the joint posterior distribution over the missing attributes are computed and the values with the highest probability are used for substituting the missing values. Computing the joint probability over multiple missing values captures the correlations between the missing values, which gives Bayes networks a clear edge over AFDs.  
In contrast, imputation using AFDs uses the AFD with the highest confidence for each attribute for prediction. If an attribute in the determining set of an AFD is missing, then that attribute is first predicted using other AFDs (chaining), before the original attribute can be predicted. The most likely value for each attribute is used for completing the missing value. When multiple missing values need to predicted, each value is predicted independently. 

We use the Cars and Adult databases described in the previous section. We compare AFD approach used in QPIAD which uses Naive Bayesian Classifiers to represent value distributions with exact and approximate inference in Bayes networks. We call exact inference in Bayes network as BN-Exact. We use Gibbs sampling as the approximate inference technique, which we call BN-Gibbs. For BN-Gibbs, the probabilities are computed using 250 samples.  

\begin{figure}
\begin{center}
\includegraphics[keepaspectratio, width=0.66\textwidth, clip, trim= 50pt 50pt 0pt 50pt]{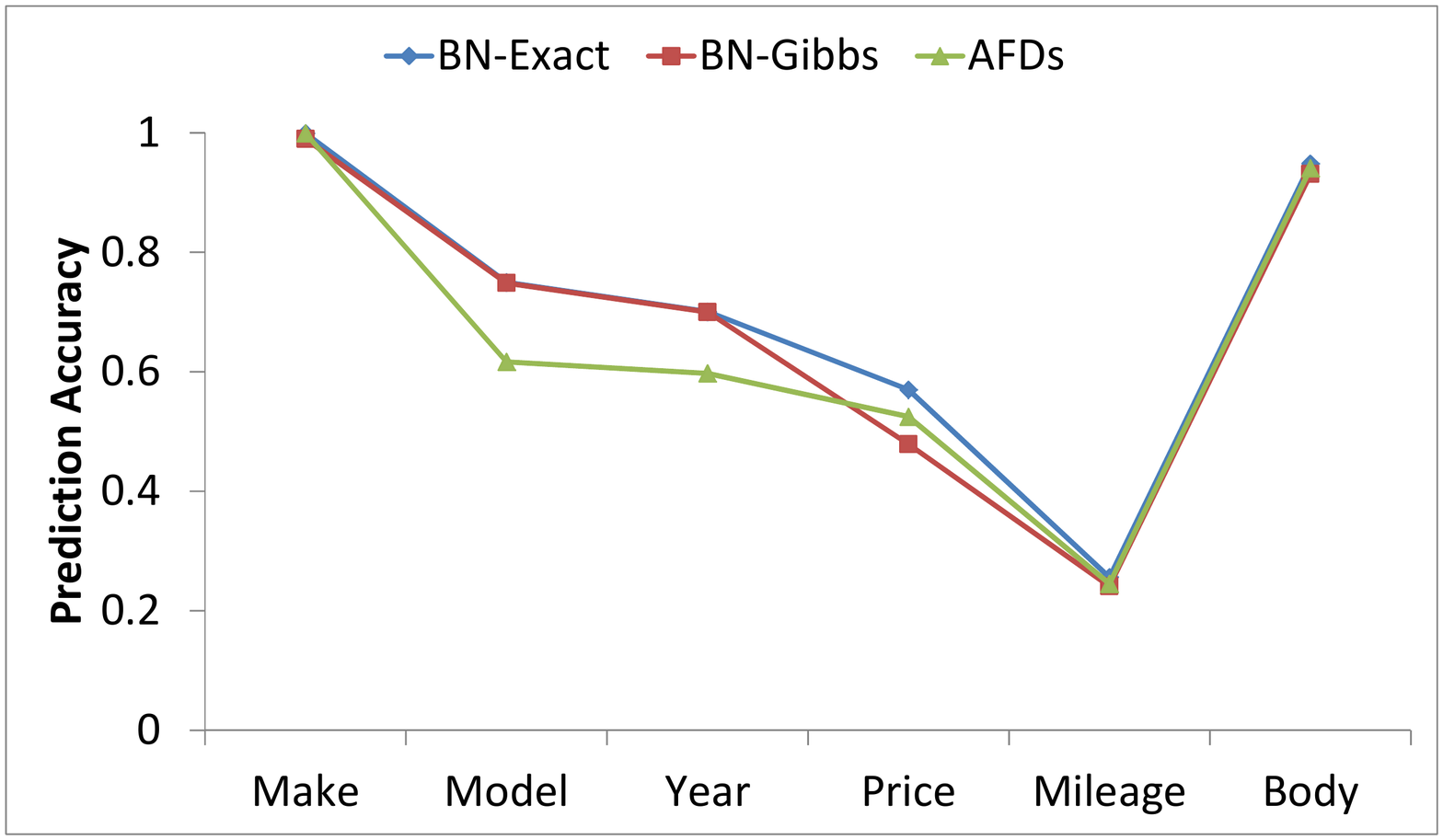}\\ 
\caption{Single Attribute Prediction Accuracy (Cars) \label{fig:impt-cars-single}}
\end{center}
\end{figure}

\medskip\mund{Imputing Single Missing Values:}
\label{subsec:impt-single}
\indent Our experiments show that prediction accuracy using Bayes nets is higher than AFDs for attributes which have multiple high confidence rules. Approaches for combining multiple rules for classification have been shown to be ineffective by Khatri~\cite{hemal-thesis}. Since there is no straightforward way for propagating beliefs using multiple AFDs, only the AFD with the highest confidence is used for propagating beliefs. This method, however, fails to take advantage of additional information that the other rules provide. Bayes networks, on the other hand, systematically combine evidences from multiple sources. Figure~\ref{fig:impt-cars-single} shows the prediction accuracy in the presence of a single missing value for each attribute in the Cars database. We notice that there is a significant difference in prediction accuracies of attributes \emph{Model} and \emph{Year}. There are multiple rules that are mined for these two attributes but using just the rule with highest confidence, ignores the influence of the other available evidence, which affects the prediction accuracy.

\begin{figure}
\begin{center}
\includegraphics[keepaspectratio, width=0.66\textwidth, clip, trim= 50pt 50pt 0pt 50pt]{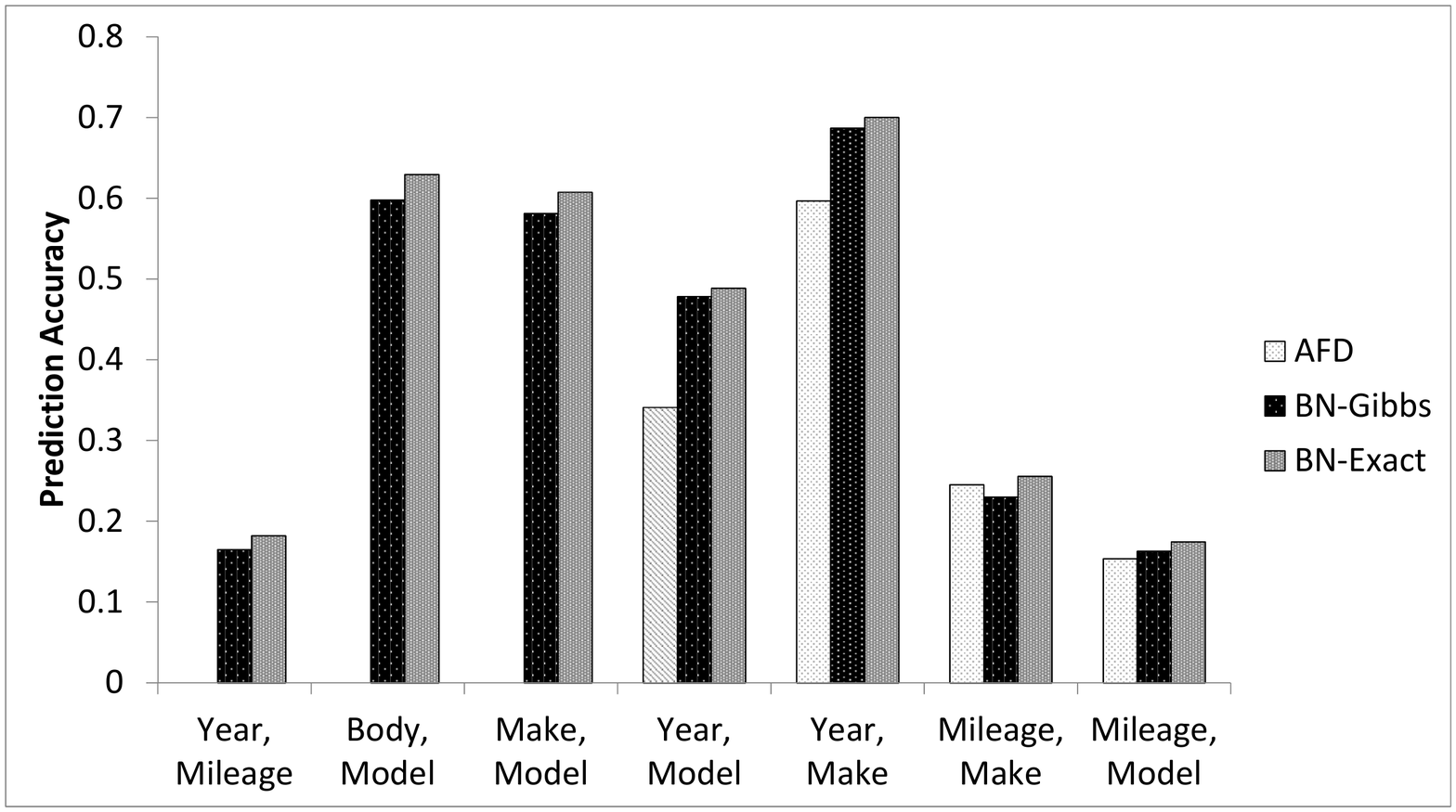}
\mvp\caption{Multiple Attribute Prediction Accuracy (Cars) \label{impt-cars-multiple}}
\end{center}
\mvp\end{figure}

\medskip\mund{Imputing Multiple Missing Values:}
\label{subsec:impt-multiple}
\indent In most real-world scenarios, however, the number of missing values per tuple is likely to be more than one. The advantage of using a more general model like Bayes networks becomes even more apparent in these cases. Firstly, AFDs cannot be used to impute all combinations of missing values, this is because when the determining set of an AFD contains a missing attribute, then the value needs to be first predicted using a different AFD by chaining. While chaining, if we come across an AFD containing the original attribute to be predicted in its determining set, then predicting the missing value becomes impossible. When the missing values are highly correlated, AFDs often get into such cyclic dependencies. In Figure~\ref{impt-cars-multiple} we can see that the attribute pairs \emph{Year-Mileage}, \emph{Body-Model} and \emph{Make-Model} cannot be predicted by AFDs. As the number of missing values increases, the number of combinations of missing values that can be predicted reduces. In our experiments with the Cars database, when predicting three missing values, only 9 out of the 20 possible combinations of missing values could be predicted. 

On the other hand, Bayes networks can predict the missing values regardless of the number and combination of values missing in a tuple. 
Secondly, while predicting the missing values, Bayes nets compute the joint probability distribution over the missing attributes which allows them to capture the correlations between the attributes. In contrast, prediction using AFDs, which use a Na\"{\i}ve Bayes Classifier to represent the value distributions, predict each of the missing attributes independently, ignoring the interactions between them. The attribute pair \emph{Year-Model} in Figure~\ref{impt-cars-multiple} shows that the prediction accuracy is significantly higher when correlations between the missing attributes are captured. We also observe that in some cases, when the missing values are D-separated~\cite{bayes-ball-paper} given the values for other attributes in the tuple, the performance of AFDs and Bayes networks is comparable. In Figure~\ref{impt-cars-multiple}, we can see that the prediction accuracy for \emph{Mileage-Make} and \emph{Mileage-Body} are comparable for all the techniques since attributes are D-separated given the other evidence in the tuple.  However, the number of attributes that are D-separated is likely to decrease with increase in incompleteness in the databases. 

\medskip\mund{Prediction Accuracy with Increase in Incompleteness in Test Data:}
\label{subsec:impt-test}
As the incompleteness in the database increases, not only does the number of values that need to be predicted increase, but also the evidence for predicting these missing values reduces. We compared the performance of Bayes nets and AFDs as the incompleteness in the autonomous databases increases. We see that the prediction accuracy of AFDs drops faster with increase in incompleteness. This is because the chaining required for predicting missing values using AFDs increases which in turn increases the chances of getting into cyclic dependencies. Also, when an attribute has multiple AFDs, propagating beliefs using just one rule and ignoring the others, often violates the principles of detachment and locality~\cite{certainty-factor-paper} impacting the prediction accuracy.

\begin{figure*}

\begin{subfigure}{0.48\textwidth}
	\includegraphics[keepaspectratio, width=\textwidth, clip, trim= 50pt 75pt 40pt 50pt]{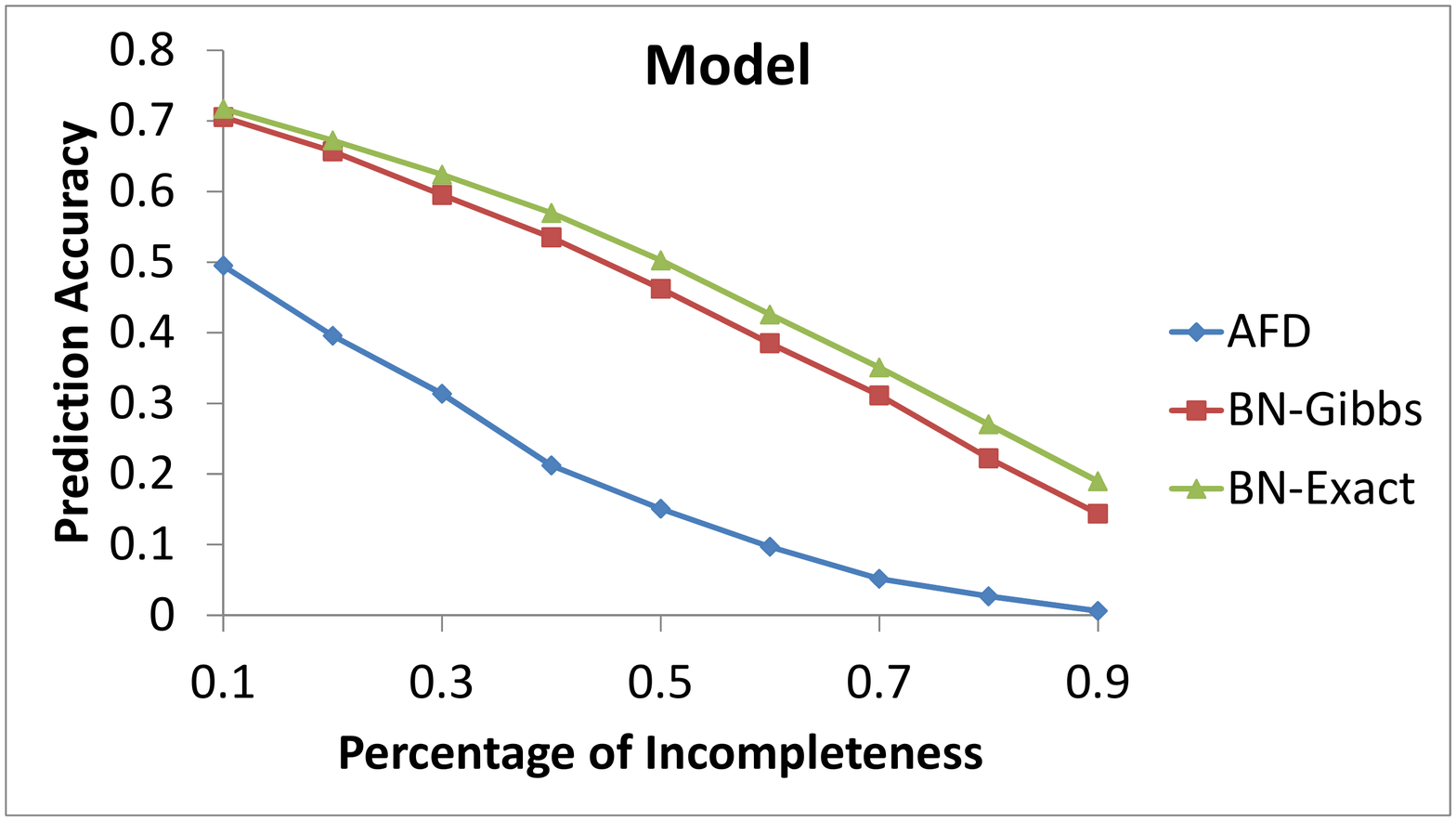}
	\caption{Model}
\end{subfigure}
\quad
\begin{subfigure}{0.48\textwidth}
	\includegraphics[keepaspectratio, width=\textwidth, clip, trim= 50pt 50pt 0pt 50pt]{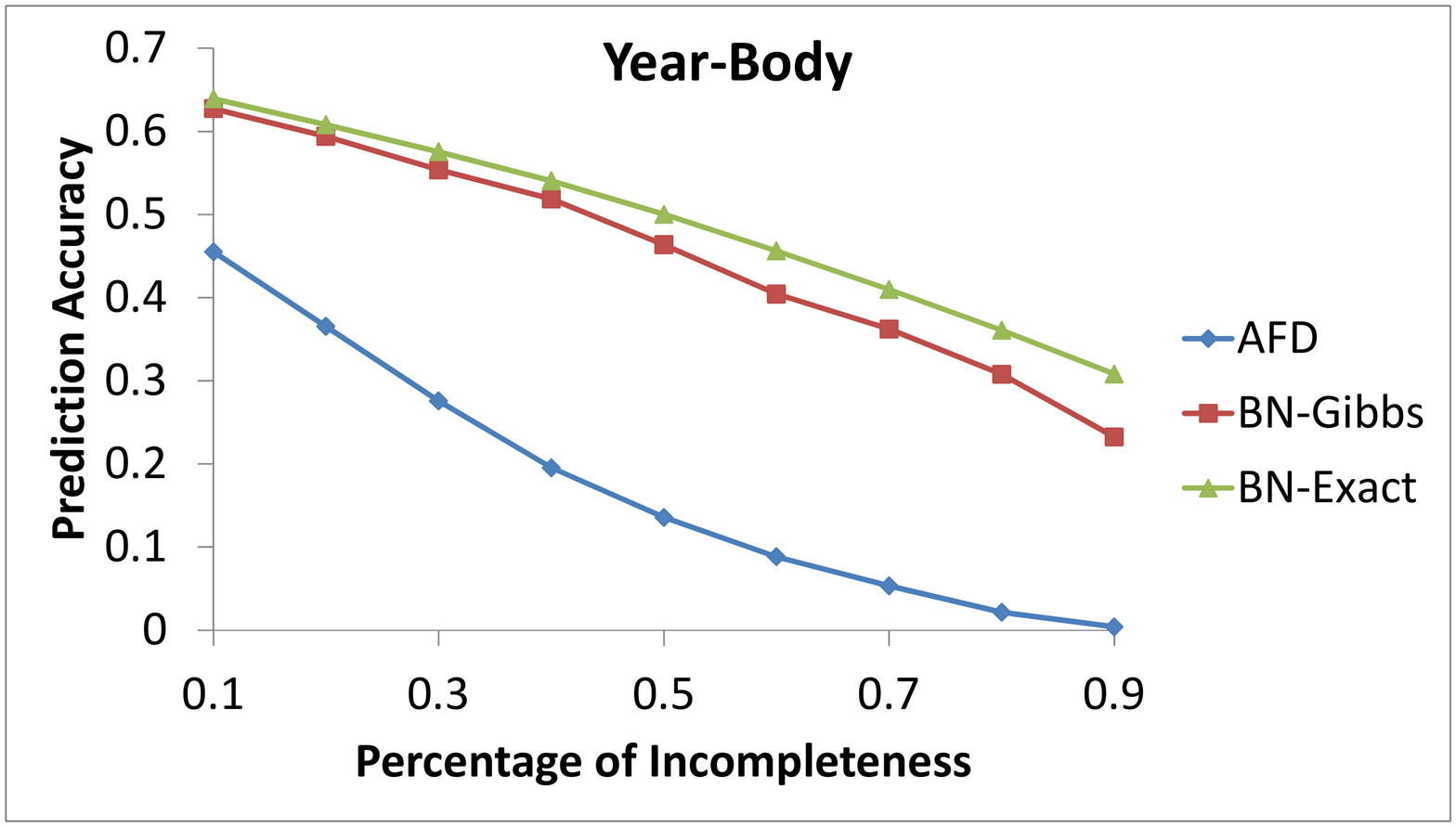}
	\caption{Year-Body}
\end{subfigure}

\begin{subfigure}{0.48\textwidth}
	\includegraphics[keepaspectratio, width=\textwidth, clip, trim= 50pt 75pt 40pt 50pt]{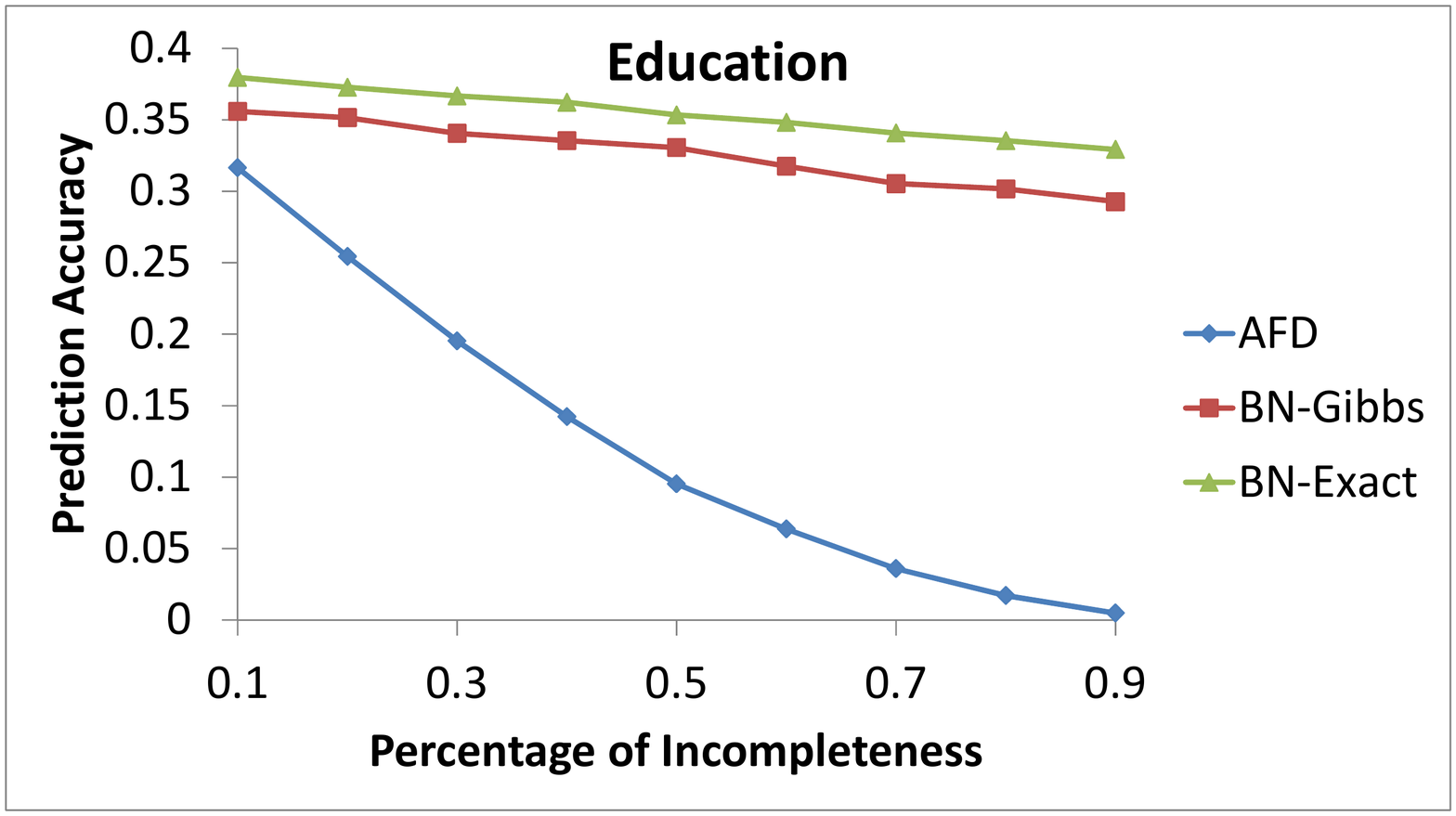}
	\caption{Education}
\end{subfigure}
\quad
\begin{subfigure}{0.48\textwidth}
	\includegraphics[keepaspectratio, width=\textwidth, clip, trim= 50pt 80pt 0pt 50pt]{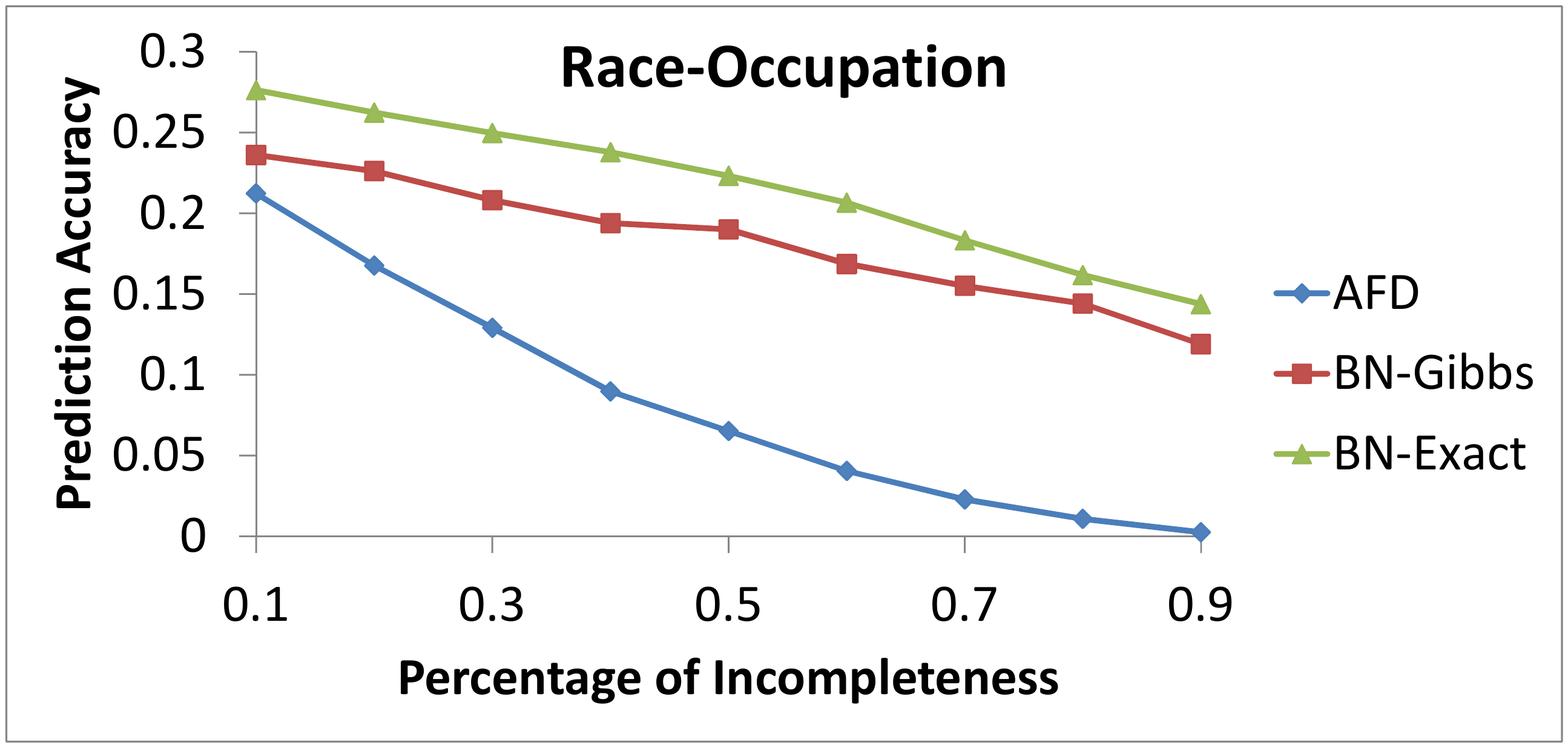}\\
	\caption{Race-Occupation}
\end{subfigure}
\caption{Prediction Accuracy with increase in percentage of incompleteness in test data for various missing attributes. (a-c) Car database, (d) Adult database.}
\label{impt-3-plots}
\end{figure*}

On the other hand, Bayesian networks, being a generative model, can systematically infer the values of any set of attributes given the evidence of any other set. Therefore, as the incompleteness of the database increases, the prediction accuracy of Bayes Networks will be significantly higher than that of AFDs. 
Figure~\ref{impt-3-plots} shows the prediction accuracy of AFDs and Bayes nets when  single and multiple attributes need to be predicted in the Cars and Adult databases. We see that both Bayes net approaches have a higher prediction accuracy than AFDs at all levels of incompleteness.

\medskip\mund{Time Taken for Imputation:}
\label{subsec:impt-cost}
We now compare the time taken for imputing the missing values using AFDs, Exact Inference (Junction tree) and Gibbs Sampling (250 samples) as the number of missing values in the autonomous database increases. Table~\ref{tb:impt-cost} reports the time taken to impute a Cars database with 5479 tuples. We note that while imputation is most accurate when using Exact Inference, the preferred method for most applications is Gibbs sampling, as it's accuracy is not very far off from exact inference (see Figure~\ref{impt-3-plots}), while keeping the cost of inference more manageable.
\begin{table}
{\footnotesize
\begin{center}
\begin{tabular}{ | p{1.6cm} | p{1.6cm} | p{1.6cm} | p{1.6cm} |}

    \hline
    Percentage of Incompleteness & Time Taken for AFD(Sec.) & Time Taken by BN-Gibbs (Sec.) & Time Taken by BN-Exact (Sec.) \\ \hline
   0\% & 0.271 & 44.46  & 16.23\\ \hline
   10\% & 0.267 & 47.15 & 44.88\\ \hline
   20\% & 0.205 & 52.02& 82.52\\ \hline
 	 30\% & 0.232 & 54.86 & 128.26\\ \hline
 40\% & 0.231 & 56.19 & 182.33\\ \hline
 50\% & 0.234 & 58.12 & 248.75\\ \hline
 60\% & 0.232 & 60.09 & 323.78\\ \hline
 70\% & 0.235 & 61.52 & 402.13\\ \hline
 80\% & 0.262 & 63.69 & 490.31\\ \hline
 90\% & 0.219 & 66.19 & 609.65\\ \hline
\end{tabular}
\caption[Time taken for Imputation by AFDs, BN-Gibbs and BN-Exact]{Time taken for predicting 5479 tuples by AFDs, BN-Gibbs (250 Samples) and BN-Exact in seconds.\label{tb:impt-cost}}
\end{center}
}
\end{table}


\section{Query Rewriting with Bayes Networks}
\label{sec:rewriting}
In information integration scenarios when the underlying data sources are autonomous, missing values cannot be completed using imputation. Our goal is to retrieve \emph{all} relevant answers to the user's query, including tuples which are relevant, but have missing values on the attributes constrained in the user's query. Since query processors are allowed read-only access to these databases, the only way to retrieve the relevant answers with missing values on query-constrained attributes is by generating and sending a set of reformulated queries that constrain other relevant attributes. We describe two techniques -- BN-All-MB and BN-Beam -- for retrieving such relevant incomplete results using Bayes networks. 

\begin{table}
{\footnotesize
\begin{center}
\begin{tabular}{ | p{0.5cm} | p{1cm} | p{1cm} | p{1cm} | p{1cm}| p{1cm}| p{1cm}|}

    \hline
    ID & Make & Model & Year & Body & Mileage \\ \hline\hline
    1 & Audi & A8 & 2005 & Sedan & 20000 \\ \hline
    2 & Audi & A8 & 2005 & null  & 15000 \\ \hline
    3 & Acura & tl & 2003 & Sedan & null \\ \hline
	 4 & BMW & 745 & 2002 & Sedan & 40000 \\ \hline
    5 & null & 745 & 2002 & Sedan & null \\ \hline
    6 & null & 645 & 1999 & Convt & null \\ \hline
    7 & null  & 645 & 1999 & Coupe & null \\ \hline
    8 & null & 645 & 1999 & Convt & null\\ \hline
    9 & BMW  & 645 & 1999 & Coupe & 40000\\ \hline
   10 & BMW & 645 & 1999 & Convt & 40000 \\ \hline  

    \end{tabular}
\end{center}
}
\mvp\caption{A Fragment of a car database}
\label{Fragment-Cars-DB}

\mvp\end{table}

\begin{algorithm}
\caption{Algorithm for BN-All-MB}
{\footnotesize
\fbox{

	\begin{minipage}[b]{3in}
 	Let \emph{R}($\text{\emph{A}}_{\text{\emph{1}}}, \text{\emph{A}}_{\text{\emph{2}}},.., \text{\emph{A}}_{\text{\emph{n}}}$) be a database relation. Suppose \emph{MB}($\text{\emph{A}}_{\text{\emph{p}}}$) is the set of attributes in the markov blanket of attribute $\text{\emph{A}}_{\text{\emph{p}}}$. A query \emph{Q}: $\sigma_{\text{\emph{A}}_{\text{\emph{p}}}=\text{\emph{v}}_{\text{\emph{p}}}}$ is processed as follows

\begin{enumerate}

\item Send \emph{Q} to the database to retrieve the base result set $\text{\emph{RS(Q)}}$. Show $\text{\emph{RS(Q)}}$, the set of certain answers, to the user. 
\item Generate a set of new queries $\text{\emph{Q'}}$, order them, and send the most relevant ones to the database to retrieve the extended result set $\widehat{\text{\emph{RS}}}(\text{\emph{Q}})$ as relevant possible answers of \emph{Q}. This step contains the following tasks.
	\begin{enumerate}
	\item \emph{Generate Rewritten Queries.}
		 Let $\pi_{(\text{\emph{MB}}(\text{\emph{A}}_{\text{\emph{p}}}))}(\text{\emph{RS(Q)}})$ be the projection of $\text{\emph{RS(Q)}}$ onto $\text{\emph{MB}}$($\text{\emph{A}}_{\text{\emph{p}}}$). For each distinct tuple $\text{\emph{t}}_\text{\emph{i}}$ in  $\pi_{(\text{\emph{MB}}(\text{\emph{A}}_{\text{\emph{p}}}))}(\text{\emph{RS(Q)}})$, create a selection query $\text{\emph{Q'}}_\text{\emph{i}}$ in the following way. For each attribute $\text{\emph{A}}_\text{\emph{x}}$ in \emph{MB}($\text{\emph{A}}_\text{\emph{p}}$), create a selection predicate  $\text{\emph{A}}_\text{\emph{x}}$=$\text{\emph{t}}_\text{\emph{i}}$.$\text{\emph{v}}_\text{\emph{x}}$. The selection predicates of $\text{\emph{Q'}}_\text{\emph{i}}$ consist of the conjunction of all these
predicates 
 \item \emph{Select the Rewritten Queries.}
		For each rewritten query   $\text{\emph{Q'}}_\text{\emph{i}}$, compute the estimated precision and estimated recall using the Bayes network as explained earlier. Then order all
 $\text{\emph{Q'}}_\text{\emph{i}}$s in order of their F-Measure scores and choose the top-\emph{K} to issue to the database.
	 \item \emph{Order the Rewritten Queries.}
		The top-\emph{K}  $\text{\emph{Q'}}_\text{\emph{i}}$s are issued to the database in the decreasing order of expected precision.
	\item \emph{Retrieve extended result set.}
		Given the ordered top-\emph{K} queries \\$\{\text{\emph{Q'}}_\text{\emph{1}}, \text{\emph{Q'}}_\text{\emph{2}},...,\text{\emph{Q'}}_\text{\emph{K}}\}$ issue them to the database and retrieve their result sets. The union of result sets $\text{\emph{RS}}(\text{\emph{Q'}}_\text{\emph{1}}), \text{\emph{RS}}(\text{\emph{Q'}}_\text{\emph{2}}),...,\text{\emph{RS}}(\text{\emph{Q'}}_\text{\emph{K}})$ is the extended result set $\widehat{\text{\emph{RS}}}(\text{\emph{Q}})$.
	\end{enumerate}

\end{enumerate}

\end{minipage}
}}
\vspace{1pt}
\end{algorithm}

\subsection{Generating Rewritten Queries}
We will use a fragment of the Car database shown in Table ~\ref{Fragment-Cars-DB} to explain our approach.
Notice that tuples $\text{\emph{t}}_{\text{\emph{2}}},\text{\emph{t}}_{\text{\emph{3}}}$ have one missing (null) value and tuples  $\text{\emph{t}}_{\text{\emph{5}}}, \text{\emph{t}}_{\text{\emph{6}}}, \text{\emph{t}}_{\text{\emph{7}}}, \text{\emph{t}}_{\text{\emph{8}}}$ have two missing values. 
To illustrate query rewriting when a single attribute is constrained in the query, consider a query(\emph{Q}) $\sigma_{\text{\emph{Body=Sedan}}}$. First, the query \emph{Q} is issued to the autonomous database and all the certain answers which correspond to tuples $\text{\emph{t}}_{\text{\emph{1}}}, \text{\emph{t}}_{\text{\emph{3}}}, \text{\emph{t}}_{\text{\emph{4}}}$  and  $\text{\emph{t}}_{\text{\emph{5}}}$ in the Table ~\ref{Fragment-Cars-DB} are retrieved. This set of certain answers forms the \emph{base result set}. However, tuple $\text{\emph{t}}_{\text{\emph{2}}}$, which has a missing value for \emph{Body} (possibly due to incomplete extraction or entry error), is likely to be relevant since the value for \emph{Body} should have been \emph{Sedan} had it not been missing. 
In order to determine the attributes and their values on which the rewritten queries need to be generated, we use the Bayesian network learnt from the sample of the autonomous database.

Using the same example, we now illustrate how rewritten queries are generated. First, the set of certain answers which form the base result set are retrieved and returned to the user. The attributes on which the new queries are reformulated consist of all attributes in the \emph{markov blanket} of the original query-constrained attribute. The markov blanket of a node in a Bayesian network consists of its parent nodes, children nodes and children's other parent nodes. We consider all attributes in the markov blanket while reformulating queries because given the values of these attributes, the original query-constrained attribute is dependent on no other attribute in the Bayesian network. From the learnt Bayesian network shown in Figure ~\ref{example-bayes-network}, the markov blanket of the attribute \emph{Body} consists of attributes \{\emph{Year}, \emph{Model}\}. The value that each of the attributes in the rewritten query can be constrained to is limited to the distinct value combinations for each attribute in the base result set. This is because, the values that the other attributes take are highly likely to be present in relevant incomplete tuples. This tremendously reduces the search effort required in generating rewritten queries, without affecting the recall too much. However, at higher levels of incompleteness, this might have a notable impact on recall, in which case, we would search over the values in the entire domain of each attribute. Continuing our example, when the query \emph{Q} is sent to the database fragment shown in Table ~\ref{Fragment-Cars-DB}, tuples $\text{\emph{t}}_{\text{\emph{1}}}$, $\text{\emph{t}}_{\text{\emph{3}}}$, $\text{\emph{t}}_{\text{\emph{4}}}$  and $\text{\emph{t}}_{\text{\emph{5}}}$ are retrieved. The values over which the search is performed for \emph{Model} is \emph{\{A8, tl, 745\}} and for \emph{Year} is \emph{\{2002, 2003, 2005\}}. 

Some of the rewritten queries that can be generated by this process
are\\
$\text{\emph{Q'}}_{\text{\emph{1}}}$~:~$\sigma_{\text{\emph{Model=A8}}\wedge
  \text{\emph{Year=2005}}}$,
$\text{\emph{Q'}}_{\text{\emph{2}}}$~:~$\sigma_{\text{\emph{Model=tl}}
  \wedge \text{\emph{Year=2003}}}$ 
and\\
 $\text{\emph{Q'}}_{\text{\emph{3}}}$~:~$\sigma_{\text{\emph{Model=745}}\wedge
  \text{\emph{Year=2002}}}$. 

Each of these queries differ in the number of results that they retrieve and the fraction of retrieved results that are relevant. An important issue here is to decide which of these queries should be issued to the autonomous database and in which order. If we are allowed to send as many queries as we want, ordering the queries in terms of their \emph{expected precision} would obviate the need for ranking the relevant possible results once they are retrieved. This is because the probability that the missing value in a tuple is exactly the value the user is looking for is the same as the expected precision of the query that retrieves the tuple. However, limits are often imposed on the number of queries that can be issued to the autonomous database. These limits could be due to network or processing resources of the autonomous data sources. Given such limits, the precision of the answers need to be carefully traded off with selectivity (the number of results returned) of the queries. One way to address this challenge is to pick the top-\emph{K} queries based on the \emph{F-measure} metric~\cite{ir-textbook}, as pointed out by Wolf et al~\cite{qpiad-paper}. F-measure is defined as the weighted harmonic mean of precision (\emph{P}) and recall (\emph{R}) measures : 
\[\frac{(1+\alpha)\cdot\emph{P}\cdot\emph{R}}{\alpha\cdot\emph{P}+\emph{R}}\]
For each rewritten query, the F-measure metric is evaluated in terms of its \emph{expected precision} and \emph{expected recall}. The latter which can be computed from the \emph{expected selectivity}. Expected precision is computed from the Bayesian network and expected selectivity is computed the same way as computed by the QPIAD system~\cite{qpiad-paper}, by issuing the query to the sample of the autonomous database. For our example, the expected precision of the rewritten query $\sigma_{\text{\emph{Model=A8}} \wedge \text{\emph{Year=2005}}}$ can be computed as the P(\emph{Body=Sedan~$|$~Model=A8 $\wedge$Year=2005}) which is evaluated by inference on the learned Bayesian network. Expected selectivity is computed as \emph{SmplSel({Q})}$\cdot$\emph{SmplRatio(R)}, where \emph{SmplSel(Q)} is the sample selectivity of the query \emph{Q} and \emph{SmplRatio({R})} is the ratio of the original database size over the size of the sample. We send queries to the original database and its sample offline and use the cardinalities of the result sets to estimate the ratio.
\smallskip 

We refer to this technique for generating rewritten queries by constraining all attributes in the markov blanket as \emph{BN-All-MB}. In section~\ref{subsec:afd-bn-all-mb} we compare the performance of \emph{BN-All-MB} and AFD approaches in retrieving uncertain relevant tuples. However, the issue with constraining all attributes in the markov blanket is that its size could be arbitrarily large. As its size increases, the number of attributes that are constrained in the rewritten queries also increase. This will reduce the throughput of the queries significantly. As we mentioned earlier, in cases where the autonomous database has a limit on the number of queries to which it will respond, we need to carefully trade off precision of the rewritten queries with their throughput. BN-All-MB and AFD approaches decide upfront the attributes to be constrained and search only over the values to which the attributes will be constrained. Both these techniques try to address this issue by using the F-measure metric to pick the top-\emph{K} queries for issuing to the database-- all of which have the same number of attributes constrained. A more effective way to trade off precision with the throughput of the rewritten queries is by making an "online" decision on the number of attributes to be constrained. We propose a technique, \emph{BN-Beam} which searches over the markov blanket of the original query-constrained attribute, and picks the best subset of the attributes with high precision and throughput.  
\subsubsection{Generating Rewritten Queries using \emph{BN-Beam}}
\label{subsec:bn-beam}
We now describe BN-Beam, our technique for generating rewritten queries which finds a subset of the attributes in the markov blanket of the query-constrained attribute with high precision and throughput.
To illustrate how rewritten queries are generated using BN-Beam, consider the same query(\emph{Q}) $\sigma_{\text{\emph{Body=Sedan}}}$. First, \emph{Q} is sent to the database to retrieve the base result set. We consider the attributes in the markov blanket of the query-constrained attribute to be the potential attributes on which the new queries will be formulated.  We call this set the \emph{candidate attribute set}.

For query \emph{Q}, the candidate attribute set consists of attributes
in the markov blanket of attribute \emph{Body} which consists of
attributes \{\emph{Model}, \emph{Year}\} for the Bayes Network in
Figure ~\ref{example-bayes-network}. Once the candidate attribute
set is determined, a beam search with a beam width, \emph{K} and
depth, \emph{L}, is performed over the distinct value combinations in
the base result set of the attributes in the candidate attribute set.
For example, when the query \emph{Q} is sent to the database fragment
shown in Table ~\ref{Fragment-Cars-DB}, tuples
$\text{\emph{t}}_{\text{\emph{1}}}$,
$\text{\emph{t}}_{\text{\emph{3}}}$,
$\text{\emph{t}}_{\text{\emph{4}}}$ and
$\text{\emph{t}}_{\text{\emph{5}}}$ are retrieved. The values over
which the search is performed for \emph{Model} is \emph{\{A8, tl,
  745\}} and for \emph{Year} is \emph{\{2002, 2003, 2005\}}.  Starting
from an empty rewritten query, the beam search is performed over
multiple levels, looking to expand the partial query at the previous
level by adding an attribute-value to it. For example, at the first
level of the search five partial rewritten queries:
$\sigma_{\text{\emph{Model=745}}}$,
$\sigma_{\text{\emph{Model=A8}}}$, $\sigma_{\text{\emph{Model=tl}}}$, $\sigma_{\text{\emph{Year=2002}}}$ and $\sigma_{\text{\emph{Year=2003}}}$ may be generated. An important issue here is to decide which of the queries should be carried over to the next level of search. Since there is a limit on the number of queries that can be issued to the autonomous database and we want to generate rewritten queries with high precision and throughput while keeping query processing costs low, we pick the top-\emph{K} queries based on the F-measure metric, as described earlier. The advantage of performing a search over both attributes and values for generating rewritten queries is that there is much more control over the throughput of the rewritten queries as we can decide how many attributes will be constrained.\\
The top-\emph{K} queries at each level are carried over to the next
level for further expansion. For example, consider query
$\sigma_{\text{\emph{Model=745}}}$ which was generated at level one.
At level two, we try to create a conjunctive query of size two by
constraining the other attributes in the candidate attribute set. Say
we try to add attribute \emph{Year}, we search over the distinct
values of \emph{Year} in the base set with attribute \emph{model}
taking the value \emph{745}. At each level \emph{i}, we will have the
top-\emph{K} queries with highest F-measure values with \emph{i} or
fewer attributes constrained. The top-$\text{\emph{K}}$ queries
generated at the Level $\text{\emph{L}}$ are sorted based on expected
precision and sent to the autonomous database in that order to
retrieve the relevant possible answers. 
\smallskip
We now describe the BN-Beam algorithm for generating rewritten queries for single-attribute queries.

\begin{algorithm}
{\footnotesize
\fbox{

\begin{minipage}[b]{3in}
\caption{Algorithm for BN-Beam}

 Let $\text{\emph{R}}(\text{\emph{A}}_\text{\emph{1}}, \text{\emph{A}}_\text{\emph{2}},.., \text{\emph{A}}_\text{\emph{n}})$ be a database relation. Suppose $\text{\emph{MB(A}}_\text{\emph{{p}}}\text{\emph{)}}$ is the set of attributes in the markov blanket of attribute $\text{\emph{A}}_\text{\emph{p}}$. All the steps in processing a query $\text{\emph{Q}}$: $\sigma_{\text{\emph{A}}_\text{\emph{p}}=\text{\emph{v}}_\text{\emph{p}}}$ is the same as described for BN-All-MB except step 2(a) and 2(d).
\begin{enumerate}

  \item[2(a)]{Generate Rewritten Queries.}
		A beam search is performed over the attributes in $\text{\emph{MB(A}}_\text{\emph{{p}}}\text{\emph{)}}$ and the value for each attribute is limited to the distinct values for each attribute in $\text{\emph{RS(Q)}}$. Starting from an empty rewritten query, a partial rewritten query ($\text{\emph{PRQ}}$) is expanded, at each level, to add an attribute-value pair from the set of attributes present in $\text{\emph{MB(A}}_\text{\emph{{p}}}\text{\emph{)}}$ but not added to the partial rewritten query already. The queries with top-$\text{\emph{K}}$ values for F-measure scores, computed from the estimated precision and estimated recall computed from the sample, are carried over to the next level of the search. The search is repeated over $\text{\emph{L}}$ levels.
 \item[2(d)]{Post-filtering.} Remove the duplicates in $\widehat{\text{\emph{RS}}}(\text{\emph{Q}})$.
\end{enumerate}

\end{minipage}
 }}
\vspace{1pt}
\end{algorithm}

In step 2(d), it is important to remove duplicates from $\widehat{\text{\emph{RS}}}(\text{\emph{Q}})$. Since rewritten queries may constrain different attributes, the same tuple might be retrieved by different rewritten queries. For example, consider two rewritten queries- $\text{\emph{Q'}}_\text{\emph{1}}$:$\sigma_{\text{\emph{Model=A8}}}$ and $\text{\emph{Q'}}_{\text{\emph{2}}}$:$\sigma_{\text{\emph{Year=2005}}}$, that can be generated at level one for the same user query $\text{\emph{Q}}$, that aims to retrieve all \emph{Sedan} cars. All \emph{A8} cars manufactured in \emph{2005} will be returned in the answer sets of both queries. Therefore, we need to remove all duplicate tuples.
\subsubsection{Handling Multi-attribute Queries}
Retrieving relevant uncertain answers for multi-attribute queries has been only superficially addressed in the QPIAD system. It attempts to retrieve only uncertain answers with missing values on any one of the multiple query-constrained attributes.  Here we describe how BN-All-MB and BN-Beam can be extended to retrieve tuples with missing values on multiple query-constrained attributes.

\textbf{BN-All-MB:}
The method described to handle single-attribute queries using BN-All-MB can be easily generalized to handle multi-attribute queries. The rewritten queries generated will constrain every attribute in the union of the markov blanket of the constrained attributes.

\textbf{BN-Beam:}
Similarly, using BN-Beam to handle multi-attribute queries is simple extension of the method described for single-attribute queries. The candidate attribute set consists of the union of the attributes in the markov blanket of each query-constrained attribute. 


To illustrate how new queries are reformulated to retrieve possibly relevant answers with multiple missing values on query-constrained attributes, consider an example query \allowbreak $\sigma_{\text{\emph{Make = BMW}}\wedge  \text{\emph{Mileage = 40000}}}$ sent to database fragment in Table~\ref{Fragment-Cars-DB}. First, this query retrieves the base result set which consists of tuples $\text{\emph{t}}_{\text{\emph{4}}}, \text{\emph{t}}_\text{\emph{9}}, \text{\emph{t}}_\text{\emph{10}}$. The set of candidate attributes on which the new queries will be formulated is obtained by the union of attributes in the markov blanket of the query-constrained attributes. For the learned Bayesian network shown in Figure ~\ref{example-bayes-network}, this set consists of \{\emph{Model, Year}\}. Once the candidate attribute set is determined, a beam search with a beam width, \emph{K}, is performed similar to the case when a single attribute is constrained. At the first level of the search some of the partial rewritten queries that can be generated are $\sigma_\text{\emph{Model=745}}$, $\sigma_{\text{\emph{Model=645}}}$ and $\sigma_\text{\emph{Year=1999}}$. The top-$\text{\emph{K}}$ queries with highest F-measure values are carried over to the next level of the search. The top-$\text{\emph{K}}$ queries generated at the Level $\text{\emph{L}}$ are sent to the autonomous database to retrieve the relevant possible answers. 

\subsection{Empirical Evaluation of Query Rewriting}

The aim of the experiments reported in this section is to compare the precision and recall of the relevant uncertain results returned by rewritten queries generated by AFDs and Bayes networks for single and multi-attribute queries. As mentioned before, we use a car database extracted from \emph{Cars.com}~\cite{cars.com} with a schema \emph{Cars(Model, Year, Body, Make, Price, Mileage)} consisting of 55,000 tuples. The second database used is \emph{Adult(WorkClass, Occupation, Education, Sex, HoursPerWeek, Race, Relationship, NativeCountry, MaritalStatus, Age)} consisting of 15,000 tuples obtained from UCI~\cite{uci-repository} data repository. These datasets are partitioned into test and training sets. In most information integration scenarios having access to sufficient training data is often very expensive as it involves sampling the autonomous data sources, which is costly. Therefore, only 15\% of the tuples are used as the training set. The training set is used for learning the topology and parameters of the Bayes network and AFDs. It is also used for estimating the expected selectivity of the rewritten queries. We use the Expectation Propagation inference algorithm~\cite{EP-paper-by-Tom-Minka} (with 10 samples) available in INFER.NET software package~\cite{infernet} for carrying inference on the Bayes network.

In order to evaluate the relevance of the answers returned, we create a copy of the test dataset which serves as the \emph{ground truth} dataset. We further partition the test data into two halves. One half is used for returning the certain answers, and in the other half all the values for the constrained attribute(s) are set to null. Note that this is an aggressive setup for evaluating our system. This is because typical databases may have less than 50\% incompleteness and even the incompleteness may not be on the query-constrained attribute(s). The tuples retrieved by the rewritten queries from the test dataset are compared with the ground truth dataset to compute precision and recall. Since the answers returned by the certain result set will be the same for all techniques, we consider only uncertain answers while computing precision and recall.

\subsubsection{Comparison of Rewritten Queries Generated by AFDs and BN-All-MB}
\label{subsec:afd-bn-all-mb}
\begin{figure*}
\centering
\begin{subfigure}{0.47\textwidth}
	\includegraphics[keepaspectratio, width=\textwidth, clip, trim= 50pt 50pt 40pt 50pt]{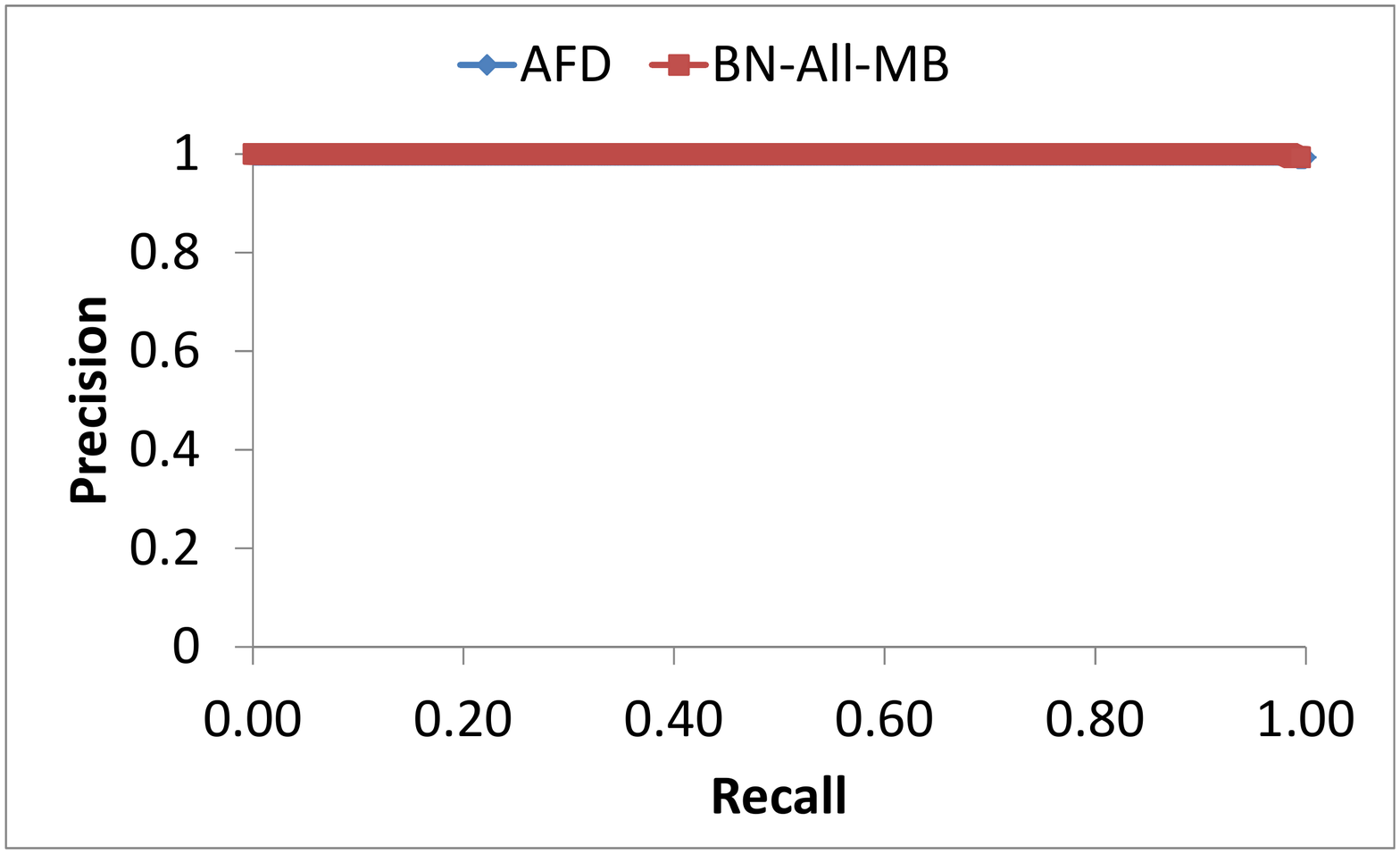}
\end{subfigure}
~
\begin{subfigure}{0.47\textwidth}
\qquad\includegraphics[keepaspectratio, width=\textwidth, clip, trim= 50pt 50pt 40pt 50pt]{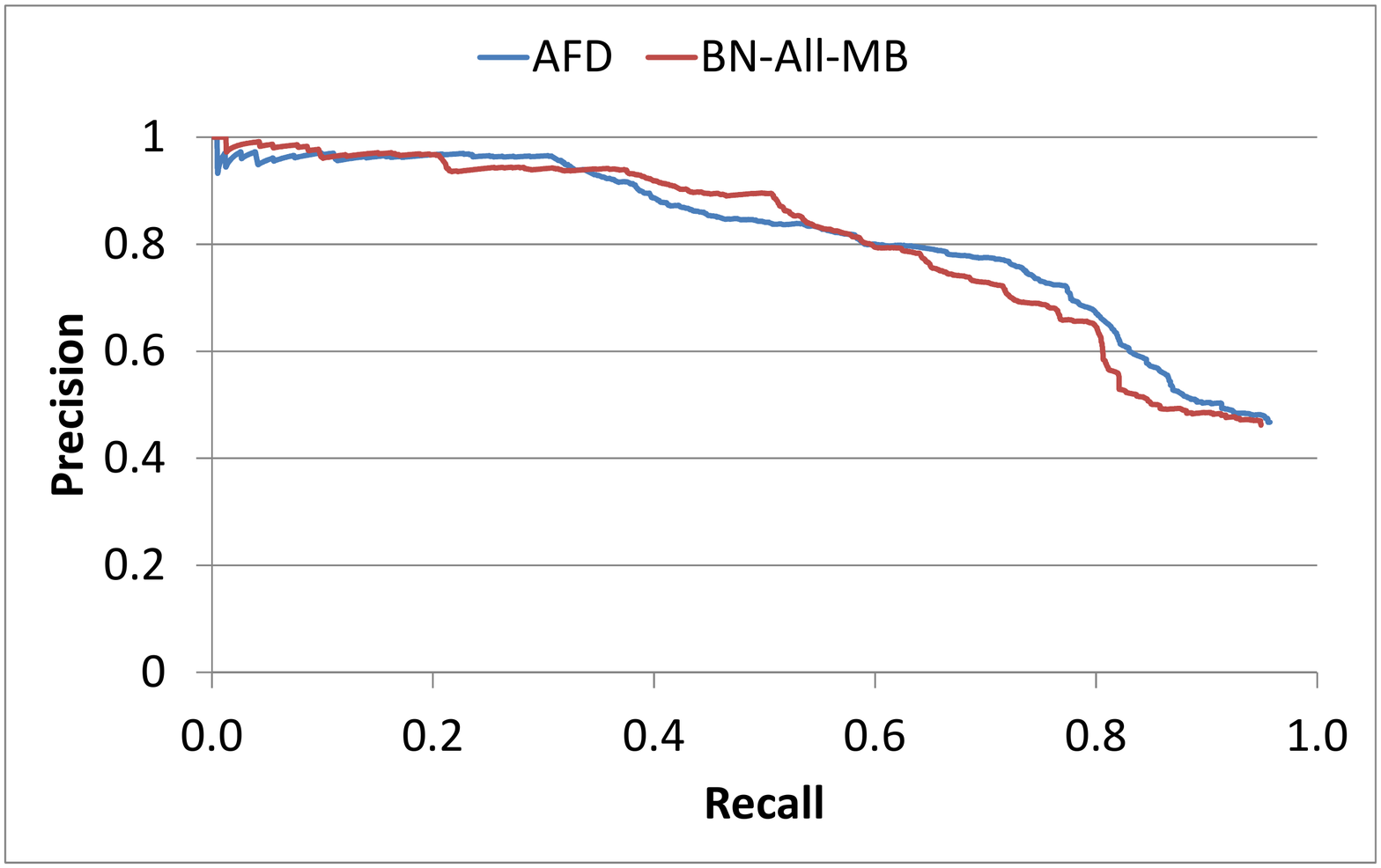}
\end{subfigure}

\begin{subfigure}{0.47\textwidth}
\includegraphics[keepaspectratio, width=\textwidth, clip, trim= 50pt 50pt 40pt 50pt]{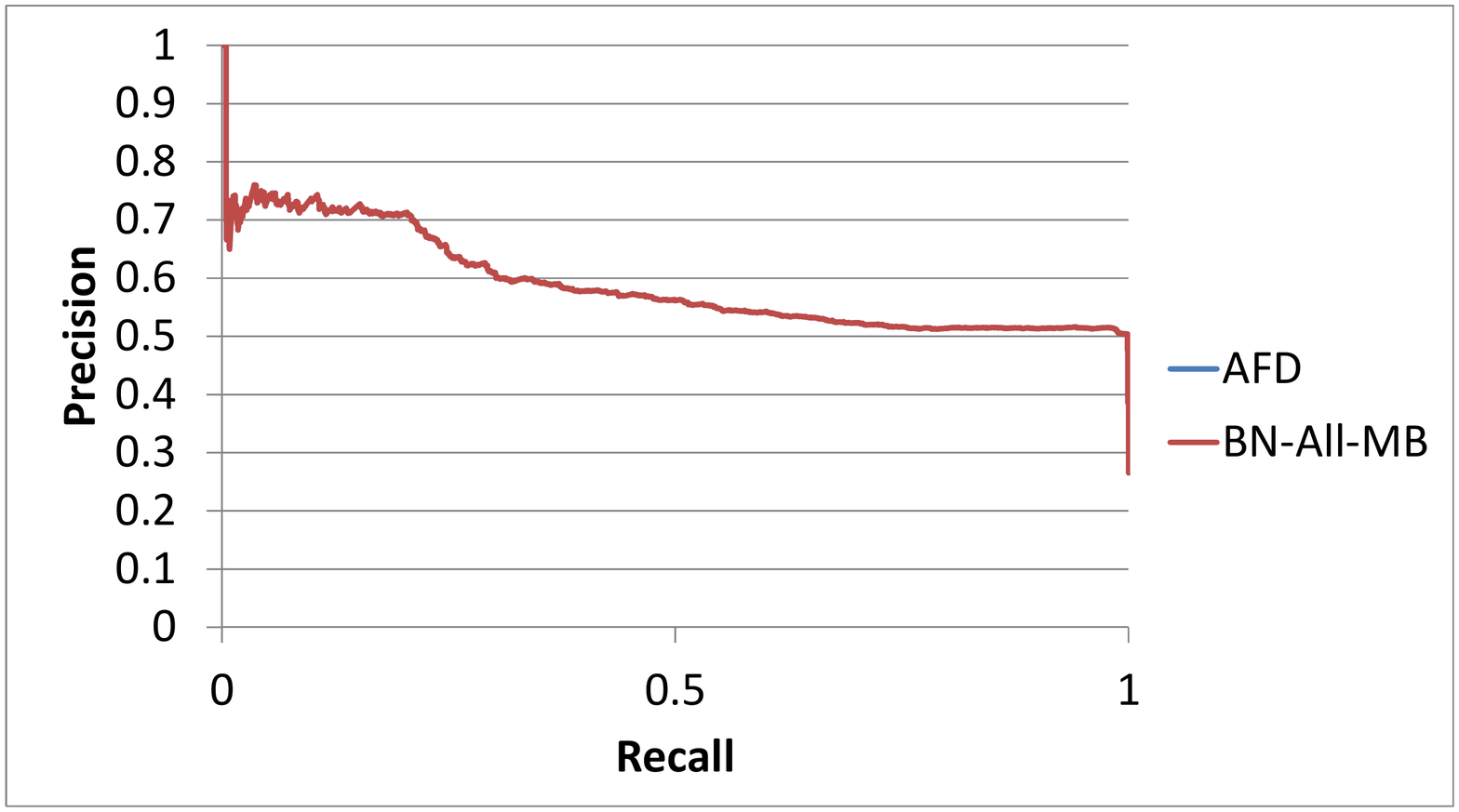}
\end{subfigure}
\mvp\caption{Precision-Recall curve for the queries
  $\sigma_{\text{\emph{Make}}}$, $\sigma_{\text{\emph{Body}}}$
 and 
$\sigma_{\text{\emph{Relationship}} = \text{\emph{Not-in-family}}}$}
  \label{fig:qr-single-precision-recall}
\end{figure*}

Figure~\ref{fig:qr-single-precision-recall} shows the precision-recall curve for queries on attribute \emph{Make} in the Car database. The size of the markov blanket and the determining set is one for attribute \emph{Make}. We note that there is no difference in the quality of the results returned by AFDs and BN-All-MB in this case (See Figure~\ref{fig:qr-single-precision-recall}). Next, we compare the quality of the results returned by AFDs and BN-All-MB when the size of the markov blanket and determining set of the AFD of the constrained attribute is greater than one. Figure~\ref{fig:qr-single-precision-recall} shows the precision and recall curves for the queries issued to Car and Adult databases. For the query on the Adult database, we found that the order in which the rewritten queries were ranked were exactly the same. Therefore, we find that the precision-recall curves of both the approaches lie one on top of the other. For the queries issued to the car database, we find that there are differences in the order in which the rewritten queries are issued to the database. However, we note that there is no clear winner. The curves lie very close to each other, alternating as the number of results returned increases. Therefore the performance of AFDs and BN-All-MB is comparable for single-attribute queries.

\subsubsection{Comparison of Rewritten Queries Generated by BN-All-MB and BN-Beam}
\label{subsec:bn-all-mb-bn-beam}
\begin{figure}
	\begin{subfigure}{0.47\textwidth}
	\includegraphics[keepaspectratio, width=\textwidth, clip, trim= 50pt 50pt 40pt 50pt]{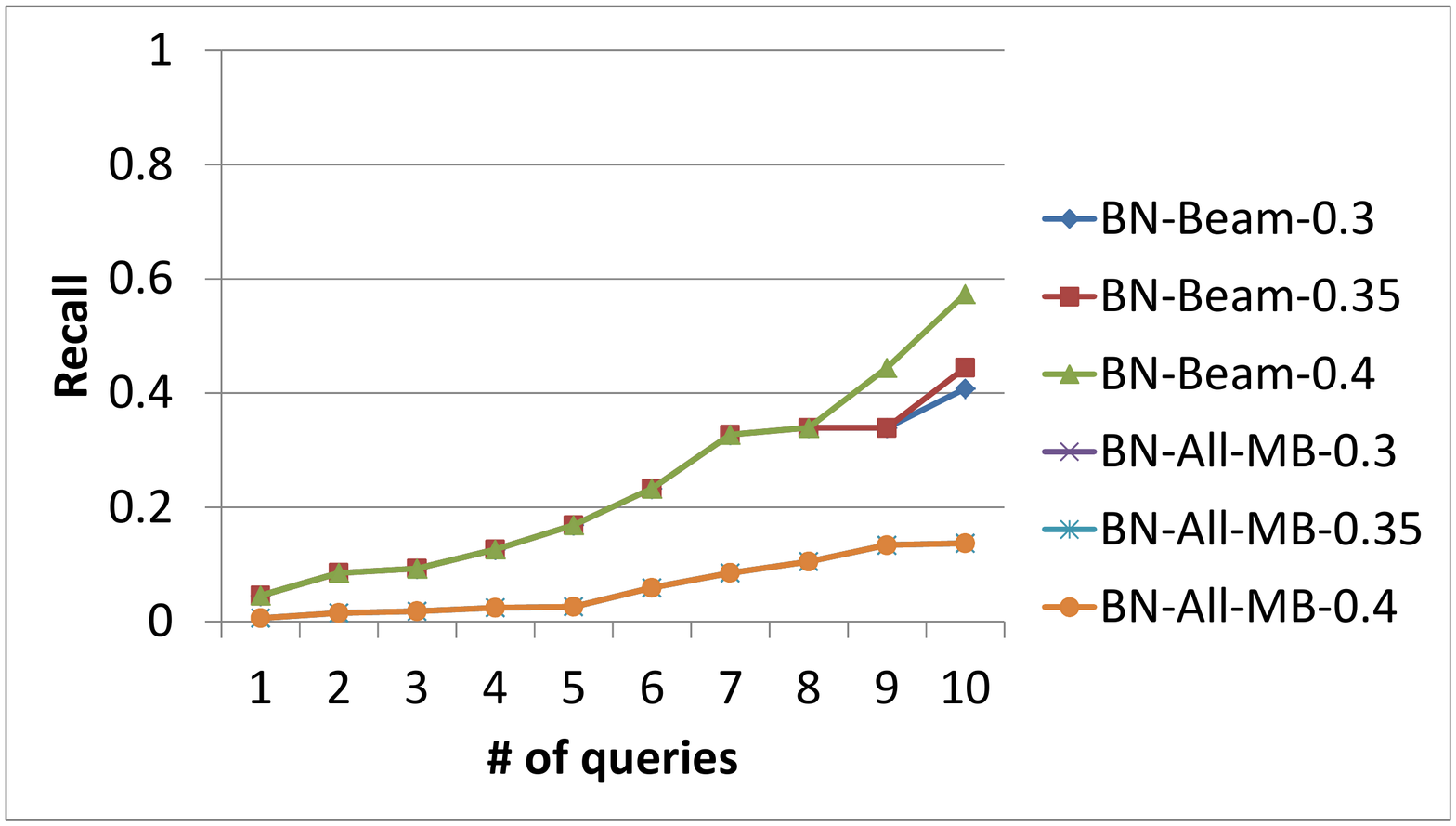}\\
	\mvp\caption{Recall}\label{fig:qr-1-myear-2002}
	\end{subfigure}
~
	\begin{subfigure}{0.47\textwidth}
	\includegraphics[keepaspectratio, width=\textwidth, clip, trim= 50pt 50pt 40pt 50pt]{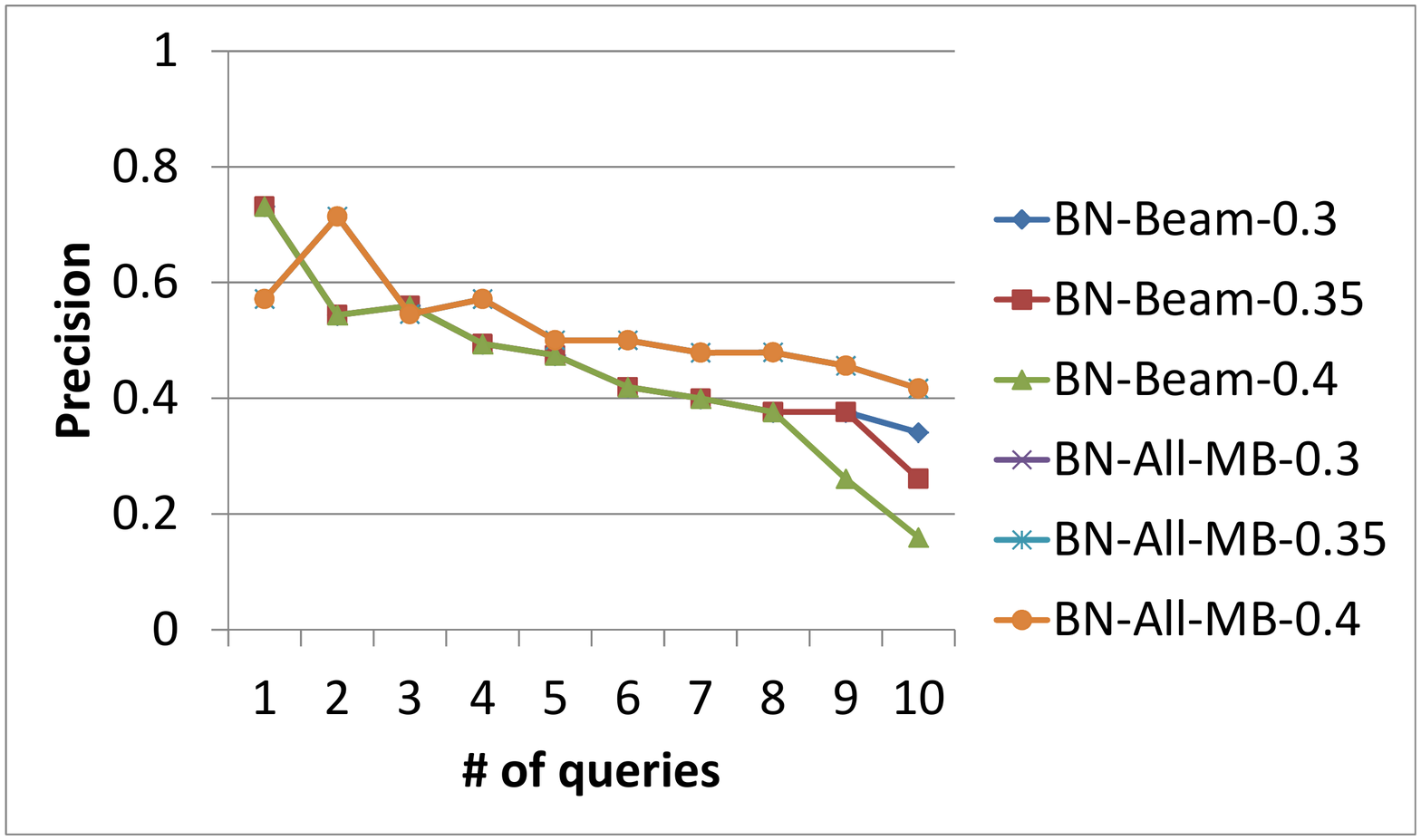}\\
	\mvp\caption{Precision}\label{fig:qr-1-myear-2002-prec}
	\end{subfigure}
\caption{Change in recall and precision for different values of $\alpha$ in F-measure metric for top-10 rewritten queries for $\sigma_{\text{\emph{Year = 2002}}}$}
\end{figure}

Figure~\ref{fig:qr-1-myear-2002} shows the increase in recall of the
results for three different values of $\alpha$ in the F-measure metric
when ten queries can be issued to the database. We refer to results
for different values for $\alpha$ for BN-Beam as BN-Beam-$\alpha$
(substitute $\alpha$ with its value) and BN-All-MB-$\alpha$ for
BN-All-MB. For BN-Beam, the level of search, $\text{\emph{L}}$, is set
to two. We note that there is no change in recall for all the three
cases for BN-All-MB. This is because there are no rewritten queries
with high throughput, therefore just increasing the $\alpha$ value
does not increase recall. For BN-Beam, we see that the recall
increases with increase in the value of $\alpha$.
Figure~\ref{fig:qr-1-myear-2002-prec} shows the change in precision
for different values of $\alpha$ as the number of queries sent to the
database increases. As expected, the precision of BN-Beam-0.30 is
higher than BN-Beam-0.35 and BN-Beam-0.40. In particular, we point out
that the precision of BN-Beam-0.30 remains competitive with
BN-All-MB-0.30 in all the cases while providing significantly higher
recall. BN-Beam is able to retrieve relevant incomplete answers
with high recall without any catastrophic decrease in precision.

\begin{figure*}
\begin{subfigure}{0.48\textwidth}
	\includegraphics[keepaspectratio, width=\textwidth, clip, trim= 60pt 100pt
	40pt 50pt]{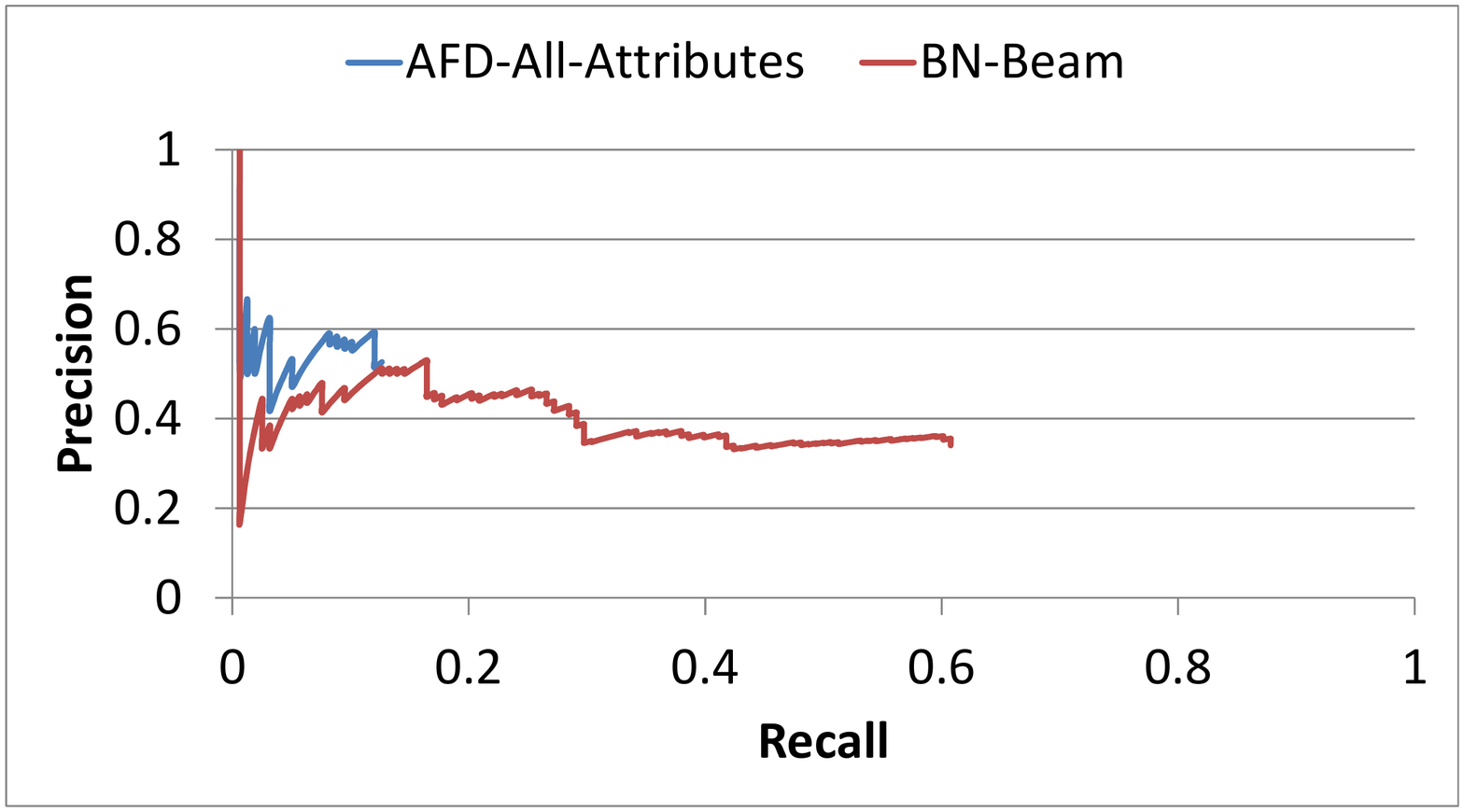}
	\mvp\caption{\large $\sigma_{\text{\emph{Make=bmw}}\wedge\text{\emph{Mileage=15000}}}$}
\end{subfigure}
~
\begin{subfigure}{0.48\textwidth}
	\includegraphics[keepaspectratio, width=\textwidth, clip, trim= 60pt 100pt 40pt 50pt]{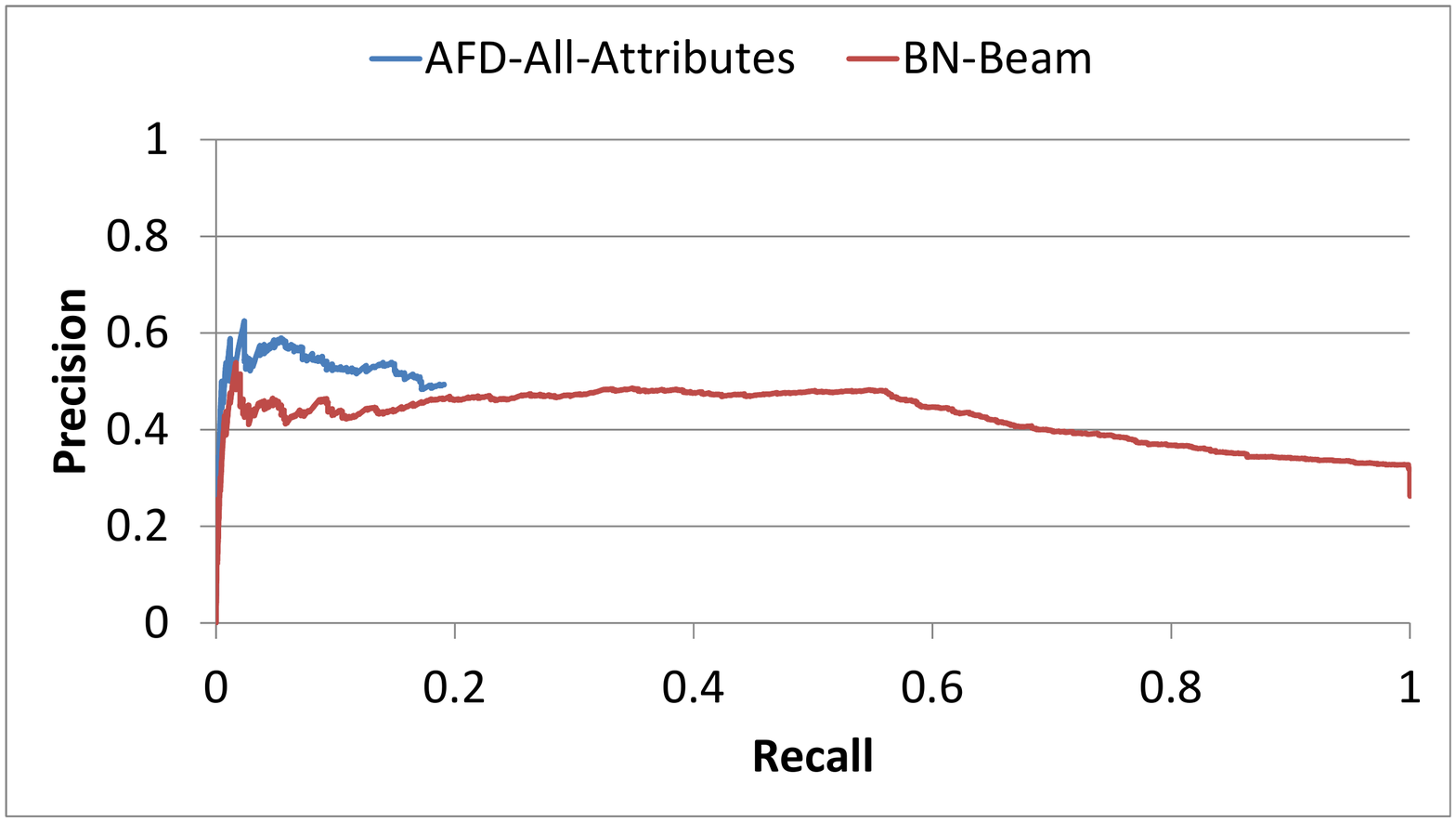}
	\mvp\caption{\large $\sigma_{\text{\emph{Education=HS-grad}}\wedge\text{\emph{Relationship=Husband}}}$}
\end{subfigure}

\begin{subfigure}{0.48\textwidth}
	\includegraphics[keepaspectratio, width=\textwidth, clip, trim= 50pt 50pt 40pt 50pt]{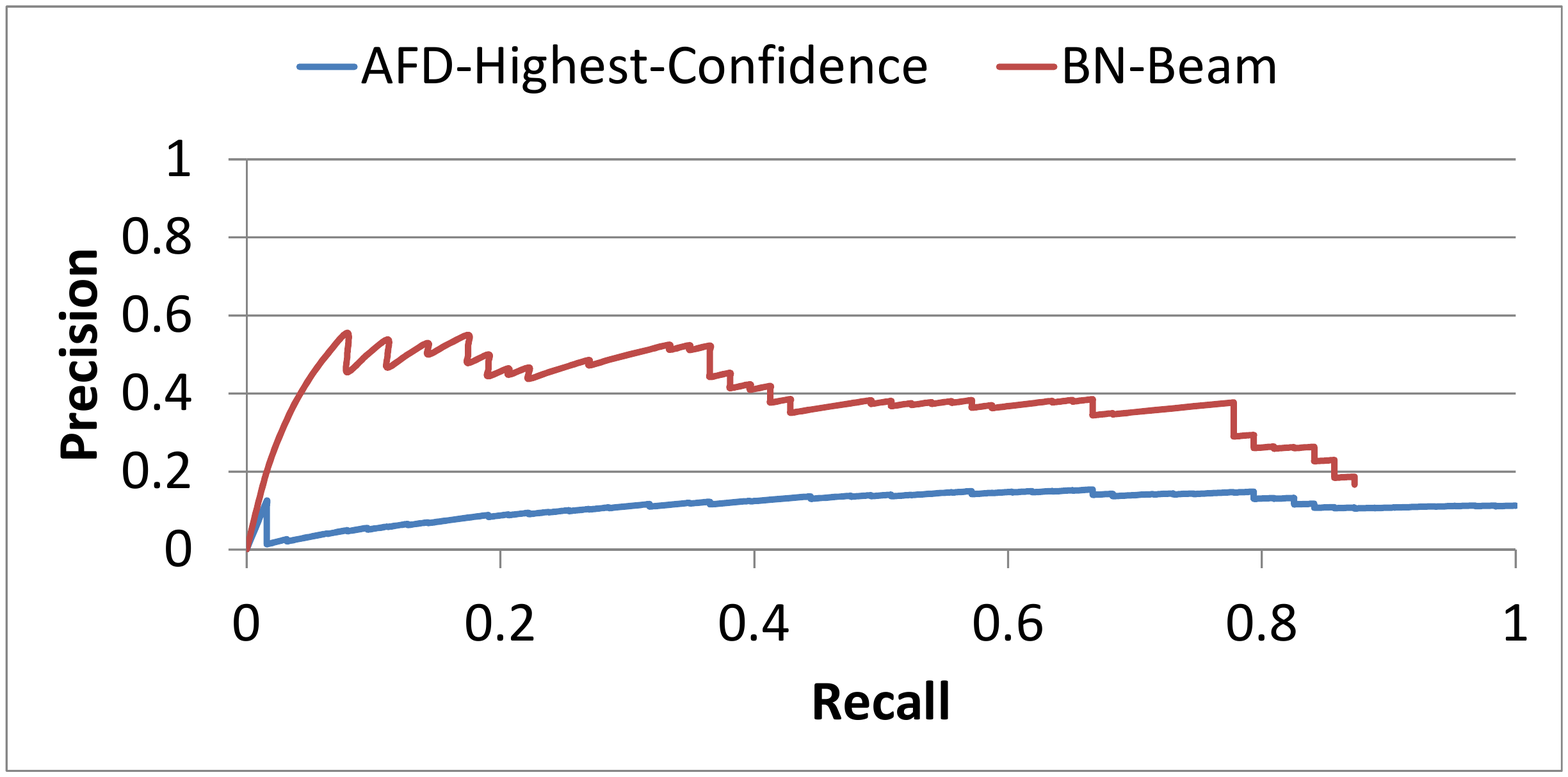}
	\mvp\caption{\large $\sigma_{\text{\emph{Make=kia}}\wedge\text{\emph{Year=2004}}}$}
\end{subfigure}
~
\begin{subfigure}{0.48\textwidth}
	\includegraphics[keepaspectratio, width=\textwidth, clip, trim= 60pt 100pt 40pt 50pt]{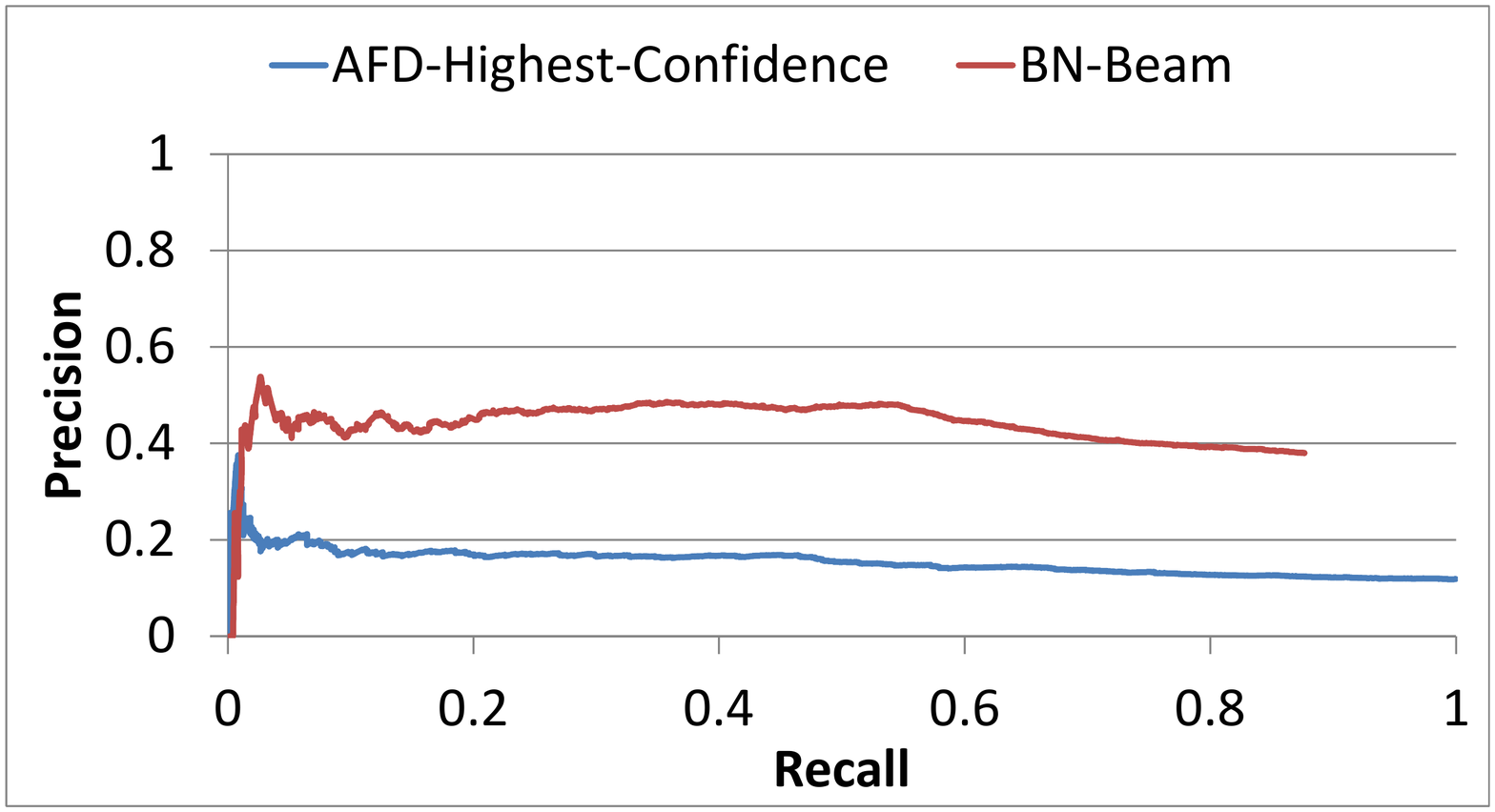}
	\mvp\caption{\large $\sigma_{\text{\emph{WorkClass=private}}\wedge\text{\emph{Relationship=own-child}}}$}
\end{subfigure}

\caption{Precision-recall curve for the results returned by top-10 rewritten queries for various queries}
\label{qr-pr-plots}
\end{figure*}

\subsubsection{Comparison of Multi-attribute Queries}
\label{subsec:afd-bn-beam}
We now compare Bayes network and AFD approaches for retrieving relevant uncertain answers with multiple missing values when multi-attribute queries are issued by the user. We note that the current QPIAD system retrieves only uncertain answers with atmost one missing value on query-constrained attributes. We compare BN-Beam with two baseline AFD approaches.\\
\textbf{1. AFD-All-Attributes:} This approach creates a conjunctive query by combining the best rewritten queries for each of the constrained attributes.  The best rewritten queries for each attribute constrained in the original query are computed independently and new rewritten queries are generated by combining a rewritten query for each of the constrained attributes. The new queries are sent to the autonomous database in the decreasing order of the product of the expected precisions of the individual rewritten queries that were combined to form the query. AFD-All-Attributes technique is only used for multi-attribute queries where the determining set of each of the attributes are disjoint.\\
\textbf{2. AFD-Highest-Confidence:} This approach uses only the AFD of the query-constrained attribute with the highest confidence for generating rewritten queries, ignoring the other attributes.\\
We evaluate these methods for selection queries with two constrained attributes. For BN-Beam, the level of search is set to 2 and the value for $\alpha$ in the F-measure metric is set to zero.\\

\subsubsection{Comparison of AFD-All-Attributes and BN-Beam}
\begin{figure*}
\begin{subfigure}{0.48\textwidth}
	\includegraphics[keepaspectratio, width=\textwidth, clip, trim= 60pt 50pt 40pt 50pt]{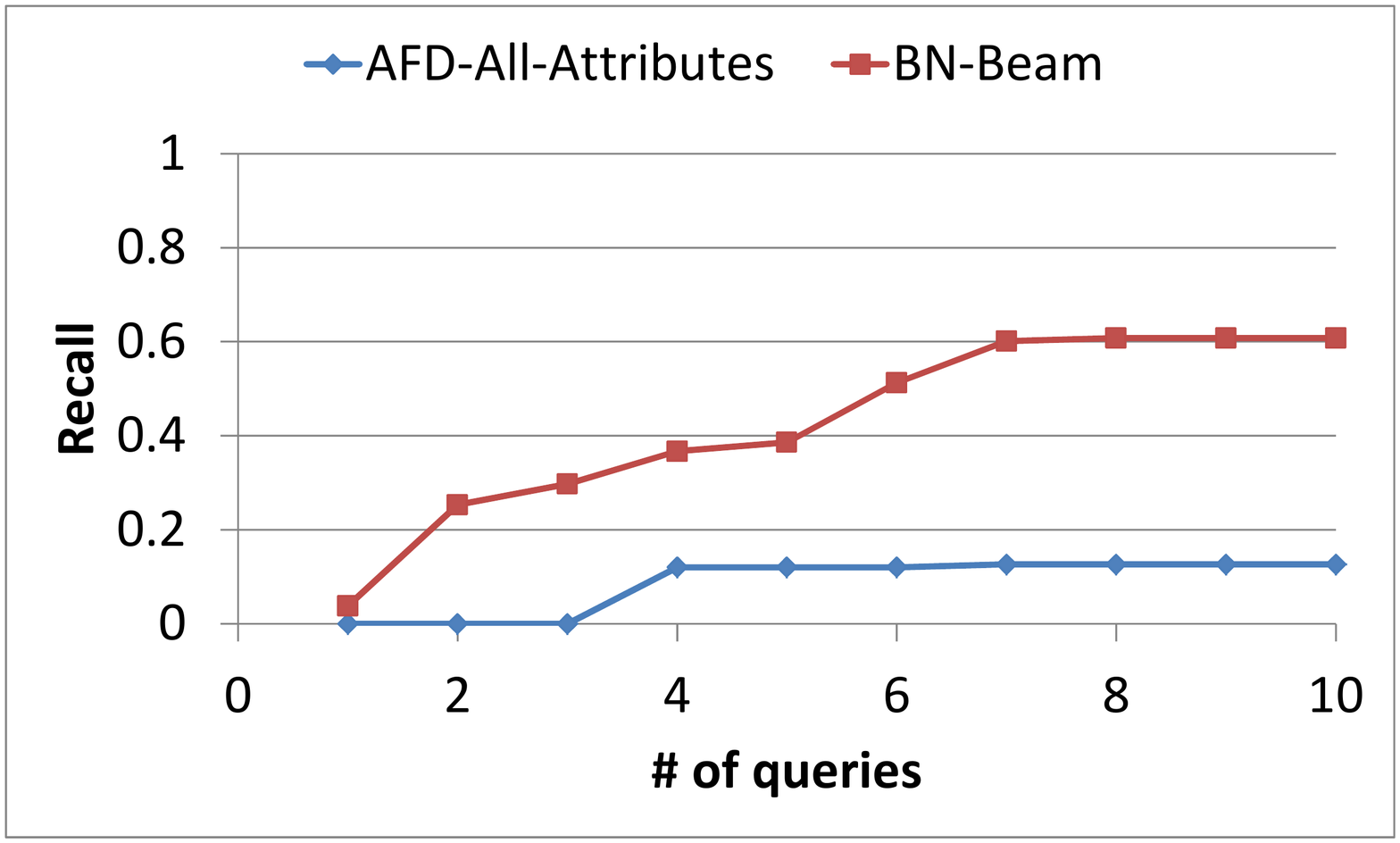}
	\mvp\caption{\large $\sigma_{\text{\emph{Make=bmw}}\wedge\text{\emph{Mileage=15000}}}$}
\end{subfigure}
~
\begin{subfigure}{0.48\textwidth}
	\includegraphics[keepaspectratio, width=\textwidth, clip, trim= 50pt 50pt 40pt 50pt]{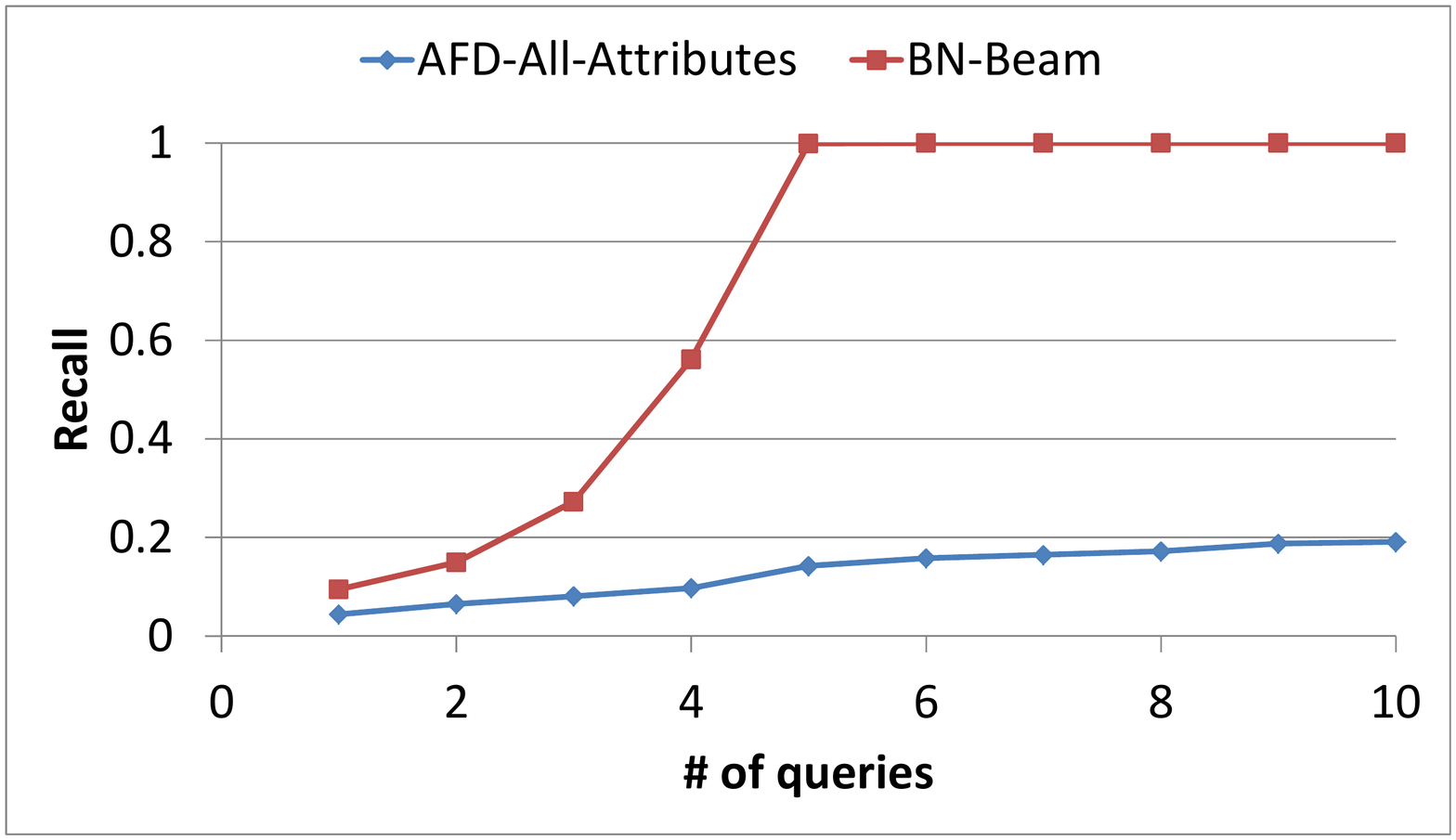}
	\mvp\caption{\large $\sigma_{\text{\emph{Education=HS-grad}}\wedge\text{\emph{Relationship=Husband}}}$}
\end{subfigure}

\begin{subfigure}{0.48\textwidth}
	\includegraphics[keepaspectratio, width=\textwidth, clip, trim= 50pt 50pt 40pt 50pt]{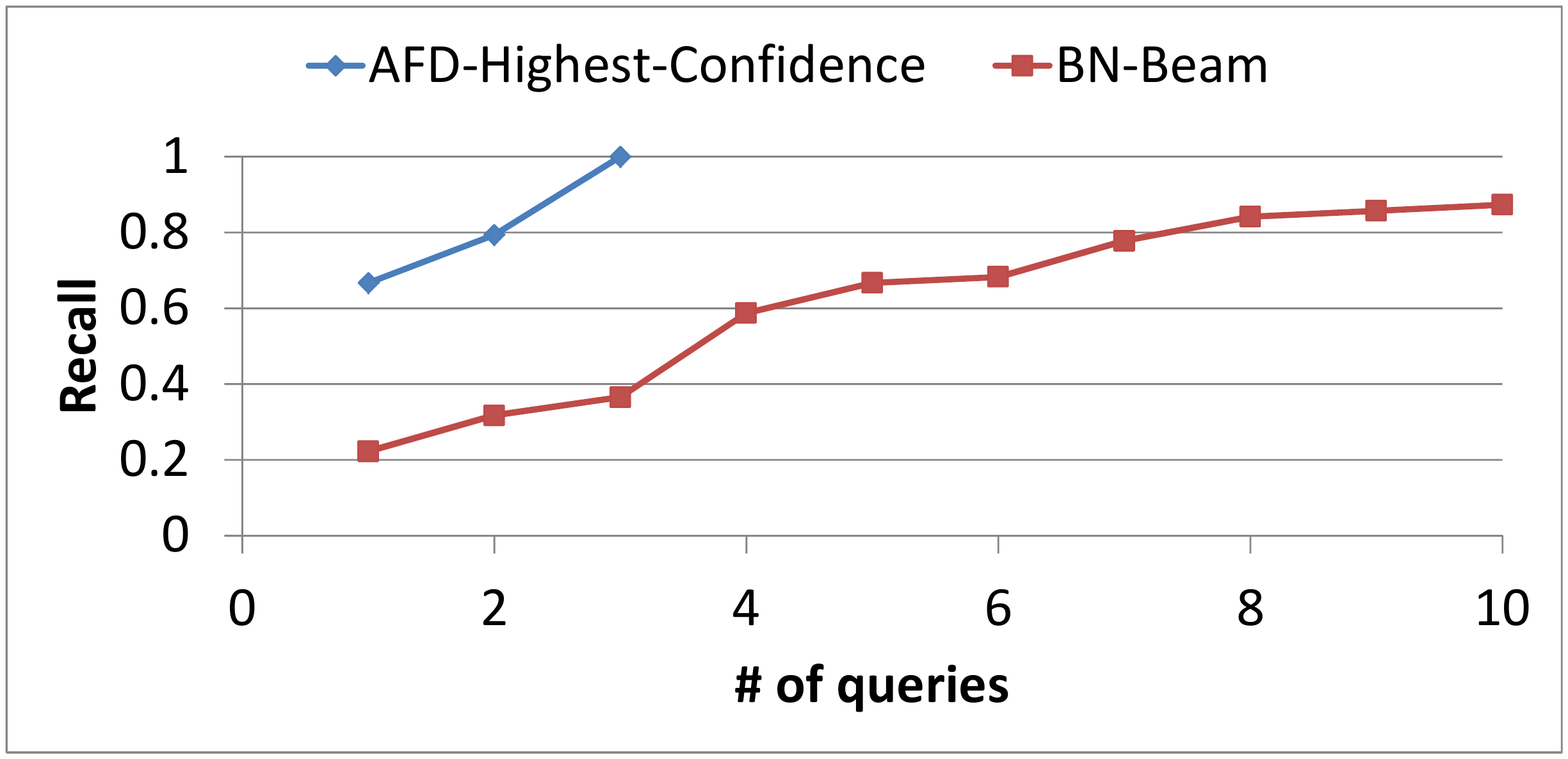}
	\mvp\caption{\large $\sigma_{\text{\emph{Make=kia}}\wedge\text{\emph{Year=2004}}}$}
\end{subfigure}
~
\begin{subfigure}{0.48\textwidth}
	\includegraphics[keepaspectratio, width=\textwidth, clip, trim= 50pt 50pt 40pt 50pt]{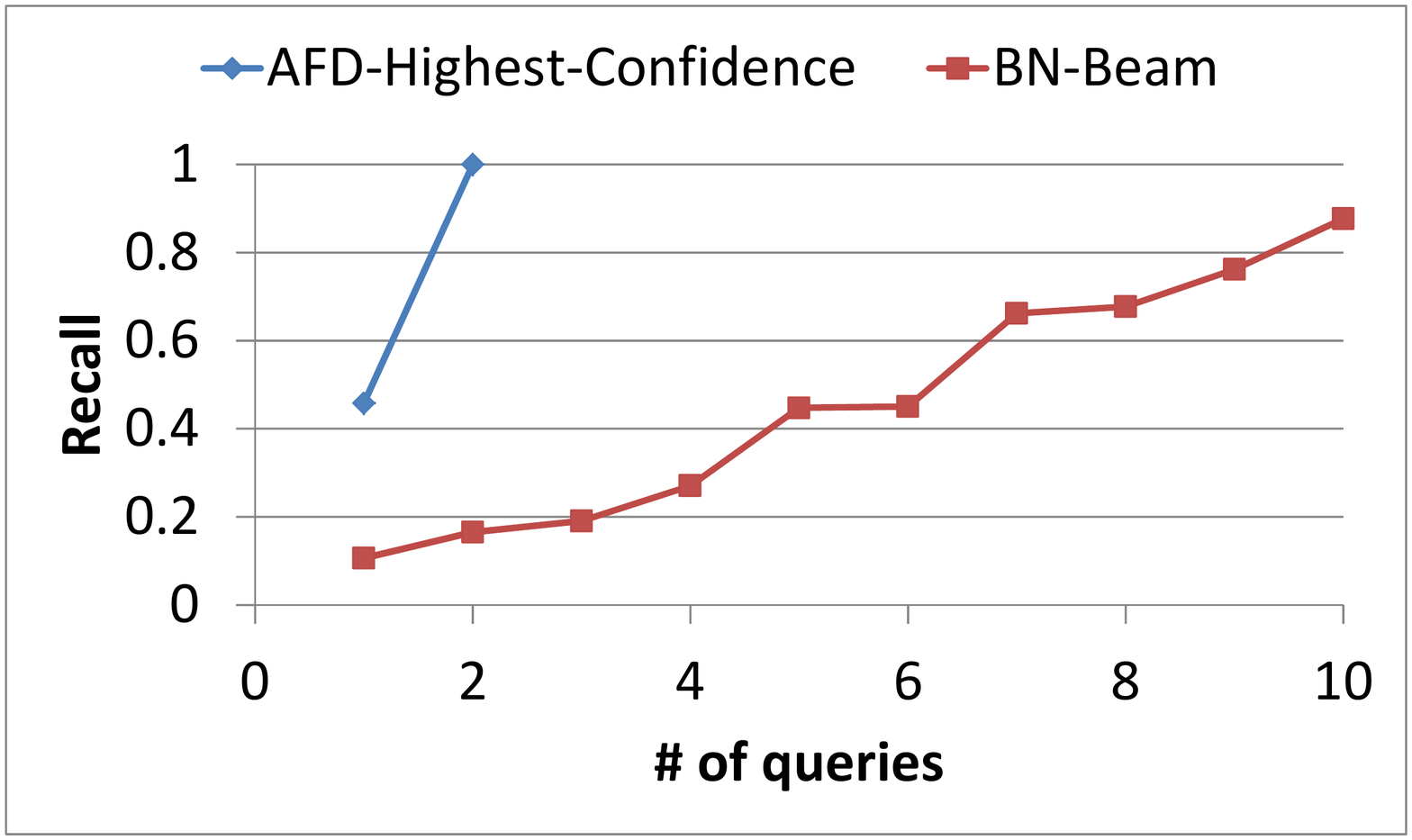}
	\mvp\caption{\large $\sigma_{\text{\emph{WorkClass=private}}\wedge\text{\emph{Relationship=own-child}}}$}
\end{subfigure}

\caption{Change in recall as the number of queries sent to the autonomous database increases for various queries}
\label{change-in-recall-plots}

\end{figure*}

Figure~\ref{qr-pr-plots} shows the precision-recall curve for the results returned by top ten rewritten queries by AFD-All-Attributes and BN-Beam for the query $\sigma_{\text{\emph{Make=bmw}} \wedge \text{\emph{Mileage=15000}}}$ issued to the Cars database. Figure~\ref{qr-pr-plots} shows a similar curve for the query \\$\sigma_{\text{\emph{Education=\allowbreak HS-grad}}\allowbreak \wedge \text{\emph{Relationship=\allowbreak Husband}}}$ issued to the Adult database. We note that the recall of the results returned by AFD-All-Attributes is significantly lower than BN-Beam in both cases (see figure~\ref{change-in-recall-plots}). This is because the new queries generated by conjoining the rewritten queries of each constrained attribute do not capture the joint distribution of the multi-attribute query. Therefore, the throughput of these queries are often very low, in the extreme case they even generate empty queries. The precision of the results returned by AFD-All-Attributes is only slightly higher than BN-Beam (See Figure~\ref{qr-pr-plots}). By retrieving answers with a little lesser precision and much higher recall than AFD-All-Attributes, BN-Beam technique becomes very effective in scenarios where the autonomous database has limits on the number of queries that it will respond to.\\

\subsubsection{Comparison of AFD-Highest-Confidence and BN-Beam}
Figure~\ref{qr-pr-plots} shows the precision-recall curves for the results returned by the top ten queries for multi-attribute queries issued to the Cars and Adult databases. Figures~\ref{change-in-recall-plots} shows the change in recall with each of the top 10 rewritten query issued to the autonomous database. We note that the recall of the results returned by AFD-Highest-Confidence is much higher than BN-Beam. However, this increase in recall is accompanied by a drastic fall in precision. This is because AFD-Highest-Confidence approach is  oblivious to the values of the other constrained attributes. Thus, this approach too, is not very effective for retrieving relevant possible answers with multiple missing values for multi-attribute queries.



\section{Conclusion}
\label{sec:conclusion}
We presented a comparison of cost and accuracy trade-offs of using Bayes network models and Approximate Functional Dependencies (AFDs) for handling incompleteness in autonomous databases. We showed how a generative model of an autonomous database can be learnt and used by query processors while keeping costs manageable.
  
We compared Bayesian networks and AFDs for imputing single and multiple missing values. We showed that Bayes networks have a significant edge over AFDs in dealing with missing values on multiple correlated attributes and at high levels of incompleteness in test data. 

Further, we presented a technique, BN-All-MB, for generating rewritten queries using Bayes networks. We then proposed a technique, BN-Beam, to generate rewritten queries that retrieve relevant uncertain results with high precision and throughput, which becomes very important when there are limits on the number of queries that autonomous databases respond to. We showed that BN-Beam trumps AFD-based approaches for handling multi-attribute queries. BN-Beam contributes to the QPIAD system by retrieving relevant uncertain answers with multiple missing values on query-constrained attributes with high precision and recall.


 
\bibliographystyle{spmpsci}
\bibliography{bibfile}

\end{document}